\documentclass[structabstract]{aa}
\usepackage{natbib}
\usepackage{amsmath}
\usepackage{graphicx}
\usepackage{txfonts}
\begin{document}

   \title{High redshift galaxies in the ALHAMBRA survey\thanks{Based on observations collected at the German-Spanish Astronomical Center, Calar Alto (CAHA), jointly operated by the Max-Planck-Institut f\"ur Astronomie (MPIA) at Heidelberg and the Instituto de Astrof\'isica de Andaluc\'ia (CSIC).}:}

   \subtitle{II. Strengthening the evidence of bright-end excess in UV luminosity functions at $2.5 \leq z \leq 4.5$ by PDF analysis}

 \author{K. Viironen
          \inst{\ref{a1}}
          \and
          C. L\'opez-Sanjuan
          \inst{\ref{a1}}
          \and
          C. Hern\'andez-Monteagudo
          \inst{\ref{a1}}
          \and
          J. Chaves-Montero
          \inst{\ref{a1}}
          B. Ascaso
          \inst{\ref{b3}}
          \and
          S. Bonoli
          \inst{\ref{a1}}
          \and
          D.~Crist\'obal-Hornillos\inst{\ref{a1}} \and
          L. A. D\'iaz-Garc\'ia
          \inst{\ref{a1}}
          \and
          A. Fern\'andez-Soto
          \inst{\ref{a12},\ref{a5}}
          \and
          I. M\'arquez
          \inst{\ref{a2}}
          \and
          J. Masegosa
          \inst{\ref{a2}}
          \and
          M. Povi\'c
          \inst{\ref{k1},\ref{a2}}
          \and
          J. Varela
          \inst{\ref{a1}}
          \and
          A. J. Cenarro
          \inst{\ref{a1}}
          \and
J.~A.~L.~Aguerri\inst{\ref{a10},\ref{a11}} \and
E.~Alfaro\inst{\ref{a2}} \and
T.~Aparicio-Villegas\inst{\ref{b1},\ref{a2}} \and
N.~Ben\'itez\inst{\ref{a2}} \and
T.~Broadhurst\inst{\ref{a6},\ref{a7}} \and
J.~Cabrera-Ca\~no\inst{\ref{a8}} \and
F.~J.~Castander\inst{\ref{a9}} \and
J.~Cepa\inst{\ref{a10},\ref{a11}} \and
M.~Cervi\~no\inst{\ref{a2},\ref{a10}} \and
R.~M.~Gonz\'alez~Delgado\inst{\ref{a2}} \and
C.~Husillos\inst{\ref{a2}} \and
L.~Infante\inst{\ref{a13},\ref{a14}} \and  
V.~J.~Mart\'inez\inst{\ref{a3},\ref{a4},\ref{a5}} \and
M.~Moles\inst{\ref{a1},\ref{a2}} \and
A.~Molino\inst{\ref{b2},\ref{a2}} \and
A.~del~Olmo\inst{\ref{a2}} \and
J.~Perea\inst{\ref{a2}} \and
F.~Prada\inst{\ref{a2}} \and 
J.~M.~Quintana\inst{\ref{a2}}
          }

\institute{Centro de Estudios de F\'isica del Cosmos de Arag\'on, Unidad Asociada al CSIC, Plaza San Juan 1, 44001 Teruel, Spain\label{a1} 
            \and
APC, AstroParticule et Cosmologie, Universit\'e Paris Diderot, CNRS/IN2P3, CEA/lrfu, Observatoire de Paris, Sorbonne Paris Cit\'e, 10, rue Alice Domon et L\'eonie Duquet, 75205 Paris Cedex 13, France\label{b3}
            \and
IAA-CSIC, Glorieta de la Astronom\'ia s/n, 18008 Granada, Spain\label{a2} 
            \and
Instituto de F\'isica de Cantabria (CSIC-UC), E-39005 Santander, Spain\label{a12}
\and
Unidad Asociada Observatorio Astron\'omico (IFCA-UV), 
            E-46980, Paterna, Spain\label{a5}
            \and
Ethiopian Space Science and Technology Institute (ESSTI), Entoto Observatory and Research Centre (EORC), Astronomy and Astrophysics Research Division, Addid Ababa, Ethiopia\label{k1}
\and
Instituto de Astrof\'isica de Canarias, V\'ia L\'actea s/n, 38200 
              La Laguna, Tenerife, Spain\label{a10}
            \and
Departamento de Astrof\'isica, Facultad de F\'isica, 
              Universidad de La Laguna, 38206 La Laguna, Spain\label{a11}            \and
Observat\'orio Nacional-MCT, Rua Jos\'e Cristino, 77. CEP 20921-400, Rio de Janeiro-RJ, Brazil\label{b1}
\and
Department of Theoretical Physics, 
            University of the Basque Country UPV/EHU, 48080 Bilbao, Spain\label{a6}
            \and
IKERBASQUE, Basque Foundation for Science, Bilbao, Spain\label{a7}
            \and
Departamento de F\'isica At\'omica, Molecular y Nuclear, 
             Facultad de F\'isica, Universidad de Sevilla, 41012 Sevilla, Spain\label{a8}
            \and
Institut de Ci\`encies de l'Espai (IEEC-CSIC), Facultat de Ci\`encies, 
             Campus UAB, 08193 Bellaterra, Spain\label{a9}
            \and
Instituto de Astrof\'{\i}sica, Universidad Cat\'olica de Chile, Av. Vicuna Mackenna 4860, 782-0436 Macul, Santiago, Chile\label{a13}
            \and
Centro de Astro-Ingenier\'{\i}a, Universidad Cat\'olica de Chile, Av. Vicuna Mackenna 4860, 782-0436 Macul, Santiago, Chile\label{a14}
            \and
Observatori Astron\`omic, Universitat de Val\`encia, 
            C/ Catedr\`atic Jos\'e Beltr\'an 2, E-46980, Paterna, Spain\label{a3}
            \and
Departament d'Astronomia i Astrof\'isica, 
            Universitat de Val\`encia, E-46100, Burjassot, Spain\label{a4}
            \and
Instituto de Astronom{\'{\i}}a, Geof{\'{\i}}sica e Ci\'encias Atmosf\'ericas, Universidade de S{\~{a}}o Paulo, S{\~{a}}o Paulo, Brazil\label{b2}
}  
         
   \date{}

 
  \abstract
   {Knowing the exact shape of the ultraviolet ($UV$) luminosity function (LF) of high-redshift galaxies is important to understand the star formation history of the early Universe. However, the uncertainties, especially at the faint and bright ends of the LFs, remain significant.} 
   {In this paper, we study the $UV$ LF of redshift $z=2.5-4.5$ galaxies in 2.38 deg$^2$ of ALHAMBRA data with $I\leq 24$. Thanks to the large area covered by ALHAMBRA, we particularly constrain the bright end of the LF. We also calculate the cosmic variance and the corresponding bias values for our sample and derive their host dark matter halo masses.}
   {We have used a novel methodology based on redshift and magnitude probability distribution functions (PDFs). This methodology robustly takes into account the uncertainties due to redshift and magnitude errors, shot noise, and cosmic variance, and models the LF in two dimensions ($z, M_{UV}$).}
   {We find an excess of bright $\sim M^{\ast}_{UV}$ galaxies as compared to the studies based on broad-band photometric data. However, our results agree well with the LF of the magnitude-selected spectroscopic VVDS data. We measure high bias values, $b\sim 8-10$, that are compatible with the previous measurements considering the redshifts and magnitudes of our galaxies and further reinforce the real high-redshift nature of our bright galaxies.}
{ We call into question the shape of the LF at its bright end; is it a double power-law as suggested by the recent broad-band photometric studies or rather a brighter Schechter function, as suggested by our multi-filter analysis and the spectroscopic VVDS data.}


   \maketitle
%

\section{Introduction}

The ultraviolet ($UV$) continuum emission of galaxies is directly proportional to their star formation rate \citep[SFR, see e.g.][]{kennicutt12}, and is conveniently observed at optical wavelengths at $z>2.5$. Hence, knowing the exact shape of the  $UV$ luminosity function (LF) at different redshifts is important to trace the star formation history of the early Universe. This, in turn, is an important piece of information to understand galaxy evolution and to constrain cosmological models.

The $UV$ LF at redshift $z\sim 2.5-4.5$ is widely studied in the literature \citep[e.g.]{steidel99,reddy09,vandenburg10,cucciati12,parsa16,mehta17}. However, the disagreement between different studies especially at the faint and bright ends of the LFs are still significant \citep[see, e.g.][]{cucciati12,parsa16}. Ideally, one should derive the LFs from a deep spectroscopic sample of galaxies without pre-selection, and at a large area. In reality this is not achievable. Selection-wise a random magnitude-limited spectroscopic survey such as the VIMOS VLT Deep Survey \citep[VVDS,][]{lefevre13} basically does the same. However, achieving both area and depth is time consuming in general, and more so in spectroscopic surveys. Hence, to our knowledge the only $z>2.5$ LF estimations based on spectroscopic data without colour pre-selection are those derived from the VVDS data \citep{paltani07,cucciati12}, while most of the high-$z$ $UV$ LFs have been derived from photometric surveys. 

Commonly, the studies of $z>2.5$ galaxies are based on samples pinpointed using the so-called drop-out technique \citep[e.g.][]{guhathakurta90,steidel96a,steidel96b,bouwens15,mehta17,ono17}. The main limitation of these studies is that the redshift distribution of the selected objects is wide and the selection is affected by significant contamination and incompleteness. The contamination can be corrected by obtaining spectroscopic redshifts \citep[see, e.g.][]{steidel96a,steidel96b,reddy06}. Incompleteness, however, is a more serious problem. As a matter of fact, it has been shown \citep{paltani07,lefevre15,inami17} that the drop-out selection leaves out a significant fraction of genuine high-$z$ galaxies. 

A selection of high redshift objects based on photometric redshifts has also been used in the literature \citep[e.g.][]{finkelstein15,parsa16}. The precision of the photometric redshifts depends strongly on the amount of filters used \citep{benitez09} and can help to tighten the redshift distributions of the selected samples. However, the above studies use the redshift information to select a list of candidates instead of directly using all the information encoded in the redshift probability distribution functions (PDFs). Hence, the problem of contamination and incompleteness remains. As a matter of fact,  \citet{viironen15} showed that in terms of colours a very conservative selection based on photometric redshifts, even when derived from various median bands leading to high-precision photo-zs, can closely resemble a drop-out selection. However, using all the information encoded in redshift PDFs allows us to exploit the colour spaces not considered in drop-out selections but which are shown to contain genuine high-z galaxies. To our knowledge the first attempt to fully exploit the redshift PDF information in LF analysis was made by \citet{mclure09}. 

In this paper we derive the $UV$ LFs for $z=2.5-4.5$ galaxies observed by the Advanced Large, Homogeneous Area Medium Band Redshift Astronomical \citep[ALHAMBRA,][]{moles08} survey. We have previously studied the galaxy number counts based on redshift PDFs in \citet{viironen15}. In this paper we derive the $UV$ LFs considering both the redshift and the $I-$band selection magnitude PDFs. The methodology used here was developed as part of the PROFUSE (PRObability Functions for Unbiased Statistical Estimations  in multi-filter surveys\footnote{http://profuse.cefca.es}) project and was introduced by \citet[][LS17 from now on]{lopez-sanjuan17} in the framework of B-band LFs of $0.2 \leq z \leq 1$ ALHAMBRA galaxies. 

The PROFUSE estimator of the LF has important  advantages with respect to previous ones. It provides a posterior two-dimensional (2D) LF at the band of interest, in our case $\Phi(z,M_{UV})$. Taking into account both the redshift and magnitude PDFs it (i) naturally accounts for $z$ and $M_{UV}$ uncertainties, (ii) ensures 100\% completeness (up to the limiting magnitude of the survey) because it works with intrinsic magnitudes instead of the observed ones, and (iii) provides a reliable covariance matrix in redshift--magnitude space. Finally, instead of modelling the LF in redshift slices, a 2D $z-M_{UV}$ model is created with the same binning as the data, and the data is fitted in two dimensions. 

The uncertainties in the derived LFs are derived considering both the shot noise and the cosmic variance. ALHAMBRA data is obtained in 48 sub-fields, allowing us to derive the latter. The cosmic variance of the galaxies is directly proportional to their bias, which in turn provides information about the mass of the hosting dark matter halo. Hence, we also derived the bias and halo masses for our sample of high-$z$ galaxies.

 The outline of this paper is as follows: Sect. \ref{sec:data} describes the ALHAMBRA data used for this study, the photometric redshifts, and the photometric pre-selection of galaxies. Sect. \ref{sec:absmag} describes the absolute magnitude estimates. Section \ref{sec:lf} describes the methodology employed to derive the LF and to model it. The corresponding error estimates are introduced in Sect. \ref{sec:errors}. The bias values and the modelling are introduced in Sect. \ref{sec:bias}. Sect. \ref{sec:results} presents the final ALHAMBRA LF and the corresponding luminosity density. In Sect.~\ref{sec:halom} the host halo masses for our sample galaxies are derived. Summary and conclusion are given in Sect.~\ref{sec:conc}.

Where necessary, we have assumed a flat $\Lambda$CDM Universe with $\Omega_{\mathrm{m}} = 0.3$, $\Omega_{\Lambda}= 0.7$, and H$_0$ = 70 km s$^{-1}$ Mpc$^{-1}$. Magnitudes are given in the AB system \citep{oke83}.

\section{The ALHAMBRA survey}\label{sec:data}

The ALHAMBRA survey \citep{moles08} has mapped a total of 4 deg$^2$ of the northern sky in eight separate fields over a seven year period ($2005-2012$). Of the total surveyed area, 2.8 deg$^2$ were completed with all the filters (2.38 deg$^2$ after masking, see Sec.~\ref{sec:sel}). ALHAMBRA uses a specially designed filter system \citep{aparicio-villegas12} which covers the optical range from 3500~\AA~to 9700~\AA~with 20 contiguous, equal width ($\sim$300~\AA~FWHM), medium-band filters, plus the three standard broad bands, $J$, $H$, and $K_s$, in the NIR. The photometric system has been specifically designed to optimise photometric redshift depth and accuracy \citep{benitez09}. The observations were carried out with the Calar Alto 3.5m telescope using two wide field cameras: LAICA in the optical, and OMEGA-2000 in the NIR. The 5$\sigma$ limiting magnitude reaches $\gtrsim24$ for all filters below 8000 \AA~and increases steeply towards redder medium band filters, up to m(AB) $\sim 21.5$ for the reddest optical filter at 9700~\AA~\citep[see Fig. 37 of][]{molino14}. In the NIR the limiting magnitudes are $J \sim 23$, $H\sim 22.5$, and $K_s\sim 22$. For details about the NIR data reduction see \citet{cristobal-hornillos09}, while the optical reduction will be described in Crist\'obal-Hornillos et al. (in prep). The ALHAMBRA catalogues and the associated Bayesian photometric redshifts are described in \citet{molino14} and are publicly available through the ALHAMBRA webpage\footnote{http://alhambrasurvey.com}. 

\subsection{ALHAMBRA photometric redshifts}\label{sec:photoz}

For all the objects in the ALHAMBRA catalogue a redshift PDF is provided as detailed in \citet{molino14}. These photometric redshifts were estimated using \texttt{BPZ2}, an updated version of the Bayesian photometric redshift (BPZ) code \citep{benitez00}. BPZ uses Bayesian inference where a maximum likelihood, resulting from a $\chi ^2$ minimisation between the observed and predicted colours for a galaxy among a range of redshifts and templates, is weighted by a prior probability. Both maximum likelihood and Bayesian redshift probability distributions are available for all the ALHAMBRA sources. The \texttt{BPZ2} spectral energy distribution (SED) library \citep[see][]{molino14} is composed of 11 SEDs: five templates for elliptical galaxies, two for spiral galaxies and four for starburst galaxies along with average emission lines and dust extinction. The opacity of the intergalactic medium has been applied as described in \citet{madau95}. The prior used gives the probability of a galaxy with apparent magnitude $m_0$ having redshift $z$ and spectral type $T$. The prior has been empirically derived for each spectral type and magnitude by fitting LFs provided by GOODS-MUSIC \citep{santini09}, COSMOS \citep{scoville07} and UDF \citep{coe06}. In \citet{viironen15} we show that the use of prior hardly affects the high redshift galaxy number counts. We checked that the high-z LFs neither are affected by the use of the prior. Following \citet{viironen15} we opted for using the maximum likelihood PDFs in the present paper.

\subsection{Photometric pre-selection}\label{sec:sel}

The photometric pre-selection was done following \citet{viironen15}. The source detections for the ALHAMBRA catalogue, consisting of 441302 objects, are made in a synthetic $F814W$ filter image, created to resemble the HST/$F814W$ filter \citep{molino14}. To avoid spurious detections, we removed the areas of low quality data, meaning those affected by bright stars or image borders, using the masks created by \citet{arnalte14}. The statistical separation between star and galaxy is encoded in the parameter Stellar\_Flag in the ALHAMBRA catalogue. We selected galaxies by setting Stellar\_Flag $\leq 0.5$. This should remove the stars up to $m<22.5$ in the reference filter, {\it F814W}. Above this magnitude the stellar flag is not defined, and slight contamination by faint stars may remain. However, for fainter magnitudes, the fraction of stars  declines rapidly compared to that of galaxies, with a contribution of $\sim10$\% for magnitudes $m(F814W) = 22.5$, declining to $\sim 1$\% for magnitudes $m(F814W) = 23.5$ \citep{molino14}. After these steps, our data consist of a total of 362788 galaxies in 2.38 deg$^2$. 

This is the only exclusive selection applied to the data. The rest of the catalogued objects are all included in the following study, but do not naturally influence the results if they either are so faint that even considering the magnitude errors they completely fall below the magnitude limit defined in Sect. \ref{sec:photom} below, or if their probability to be at the redshift range of our interest is null.

\section{Absolute magnitude estimates}\label{sec:absmag}

In this section, we follow the methodology presented in LS17. These authors calculate the LF in a synthetic $B$-band filter. In this paper, we have created a synthetic 100~\AA~wide top-hat filter centred at 1500~\AA~restframe. First we introduce in Sect. \ref{sec:photom} the $I$-band selection taking into account the magnitude errors and the correction for Eddington bias. The photometric redshift PDFs are taken into account in Sect. \ref{sec:photoz}. Finally, with all these ingredients, the rest-frame UV absolute magnitudes are introduced in Sect. \ref{sec:uvmags}.

\subsection{$I$-band selection}\label{sec:photom}

The ALHAMBRA catalogue is selected in the synthetic $F814W$ images, which we refer to here as $I$ band. Hence, all the studies based on this catalogue are affected by this selection. For each galaxy, we observe its magnitude $I$ with an error $\sigma_I$. Hence, the real magnitude, $I_0$, of the galaxy can be described as a Gaussian (in flux space) centred in $I$ and with a standard deviation $\sigma_I$ (for the magnitude space description, see Eq. [3] in LS17). In addition, we know that the faint galaxies are more numerous than the bright ones. For this reason, the net effect of photometric errors is to slightly increase the number of bright galaxies at the expense of fainter ones, that is, to flatten the increasing trend of the number counts. This effect is generally known as Eddington bias \citep{eddington1913,teerikorpi04}. To take into account these two factors, we describe each object in our catalogue by a posterior PDF:

\begin{equation}
\mathrm{PDF}(I_0|I,\sigma_I)\propto C(I_0)P(I|I_0,\sigma_I),
\end{equation}

\noindent where the integral of PDF($I_0|I,\sigma_I$) is normalised to one and the number count term $C(I_0)$ is obtained by fitting the intrinsic ($I_0$) galaxy number counts by an equation:

\begin{equation}
\log_{10}[C(I_0)] \propto -0.015I_0^2+1.00I_0.
\end{equation}

\noindent Here the intrinsic ($I_0$) number count distribution was deconvolved from the observed ($I, \sigma_I$) distribution using \texttt{emcee} \citep{foreman-mackey13} code, a Python implementation of the affine-invariant ensemble sampler for Markov chain Monte Carlo (MCMC) proposed by \citet{goodman10}.

The photometric error, $\sigma_I$, accounts for three terms:

\begin{equation}
\sigma_I = \sqrt{\sigma _{\mathrm{phot}}^2 + \sigma _{\mathrm{ZP}}^2 + \sigma _{\mathrm{sky}}^2},
\end{equation}

\noindent where $\sigma _{\mathrm{phot}}$ is the photon counting error, $\sigma _{\mathrm{ZP}}=0.02$ is the uncertainty in the zero point, and $\sigma _{\mathrm{sky}}$ the sky background uncertainty. The last was estimated empirically by placing random apertures across the empty areas of the ALHAMBRA images \citep{molino14}. 

With all these ingredients, the $I-$band selection is robustly taken into account by describing each galaxy with a source function:

\begin{equation}\label{eq:source}
S(I_0|I,\sigma_I) = \frac{1}{f_c(I_0)}\mathrm{PDF}(I_0|I,\sigma_I)\int P(I|I_0,\sigma_I)\mathrm{d}I.
\end{equation}

\noindent In this equation $f_c(I_0)$ is the completeness calculated by injecting sources of known $I_0$ magnitude in the $I$-band ALHAMBRA images and computing their detection rate (see LS17 for details), and the last term, $\int P(I|I_0,\sigma_I)\mathrm{d}I $ provides the probability that the object has a positive flux.

The final selection is then made by setting $I_0<24$, where a completeness of $f_c=0.85$ is reached on average in the 48 ALHAMBRA sub-fields (LS17). Formally, the selection is made by excluding the galaxies fulfilling the criterion

\begin{equation}
\int _{-\infty}^{24}S(I_0|I,\sigma_I)\mathrm{d}I=0.
\end{equation}

\noindent This means that even galaxies with observed magnitude $I>24$ may partially enter the selection. As emphasised in LS17, this kind of selection in real magnitudes (instead of in the observed ones which is a normal practice in the literature) assures, once the completeness correction of Eq. (\ref{eq:source}) is applied, a 100\% complete sample, with a well controlled selection function.

\subsection{Photometric redshift information}\label{sec:photoz}

The reliability of the ALHAMBRA photometric redshifts and redshift probability distribution functions are well demonstrated \citep[e.g.][]{molino14,lopez-sanjuan15b}. Further following LS17, and along the line taken in several works related to ALHAMBRA \citep{viironen15,ascaso15,lopez-sanjuan15,diaz-garcia15}, the redshifts of the galaxies are taken into account by considering their whole redshift probability distributions functions. The probability that a galaxy $i$ is located at redshift $z$ and has a spectral type $T$ is PDF$_i(z,T)$. Hence, the probability that the galaxy $i$ is located at redshift $z$ is: 

\begin{equation}
\mathrm{PDF}_i(z)=\int \mathrm{PDF}_i(z,T)\mathrm{d}T.
\end{equation}

The integral of the PDF$_i(z$) is normalised to one by definition, that is:

\begin{equation}
1=\int \mathrm{PDF}_i(z)\mathrm{d}z=\int \int \mathrm{PDF}_i(z,T)\mathrm{d}T\mathrm{d}z.
\end{equation}

\begin{figure}
\centering
\includegraphics{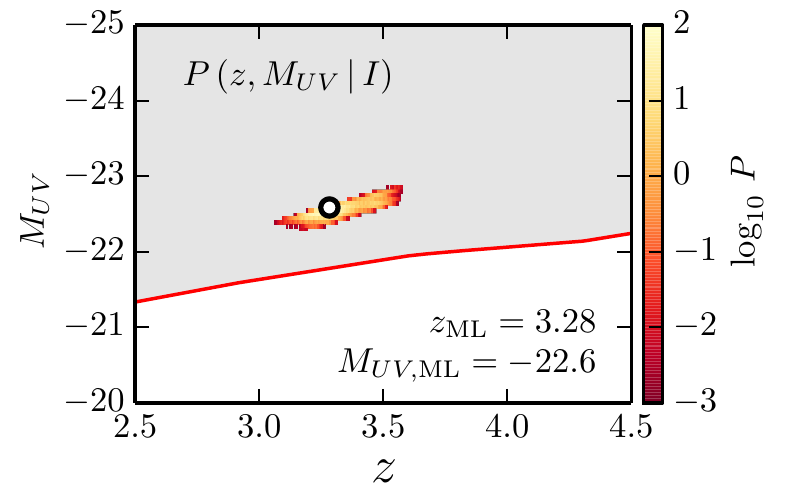}
\includegraphics{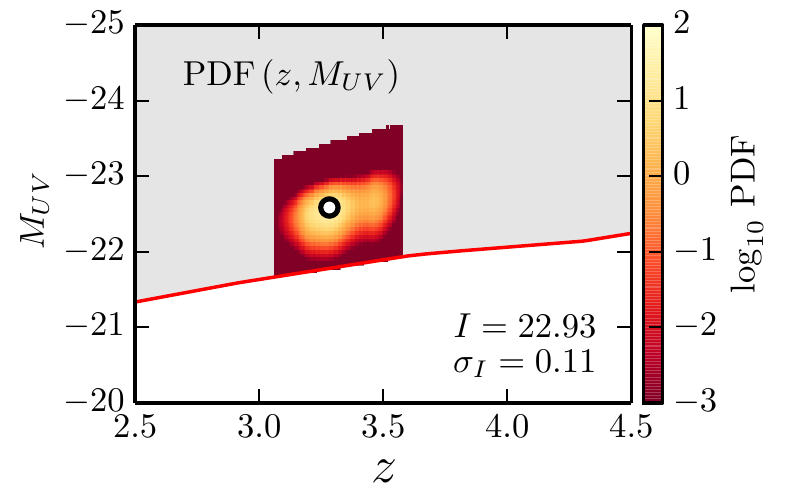}
\caption{{\it Top:} Probability of an arbitrary high-redshift ALHAMBRA galaxy with observed magnitude $I=22.93\pm 0.11$ in $z - M_{UV}$ space. {\it Bottom:} The above function convolved with the source function, $S(I_0|I,\sigma _I)$, gives the probability in $z - M_{UV}$ space for the real magnitude $I_0$. In both figures the white dot marks the maximum likelihood redshift, $z_{\mathrm{ML}}$, and the corresponding $M_{UV,\mathrm{ML}}$, labelled in the panel. The red line indicates the applied $I_0=24$ limiting magnitude, and the accessible total volume is shown as shaded grey area.}
\label{fig:PI}
\end{figure}

\subsection{$UV$ magnitudes}\label{sec:uvmags}

Armed with the source function, and bi-dimensional redshift probability distribution function, PDF($z,T$), for each galaxy we can now obtain its magnitude, as a function of both redshift and template, in the target $UV$ filter by the equation:

\begin{equation}
M_{UV}(z,T|I_0) = I_0-5\log _{10}[\mathrm{D}_{\mathrm{L}}(z)]-k(z,T)-25,
\label{eq:mabs}
\end{equation}

\noindent where D$_\mathrm{L}(z)$ is the luminosity distance in Mpc and $k(z,T)$ accounts for the $k$-correction between the observed $I$-band at redshift $z$ and the targeted $UV$-filter at rest-frame for each template. Appendix A in LS17 gives the details of how to derive the $k$-correction for the rest frame $B$-band targeted in that study. In our case the methodology is exactly the same with the only difference being that we have created our own target $UV$-filter: a top-hat filter centred at rest-frame 1500 \AA~with a width of 100 \AA.

Next, for each galaxy $i$, we constructed the probability $P_i(z,M_{UV}|I_i)$ by a PDF-weighted histogram of $M_{{UV},i}=M_{UV}(z,T|I_i)$ with a very fine binning ($dM_{UV}=0.02$):

\begin{equation}
P_i(z,M_{UV}|I_i)\mathrm{d}M_{UV}=\int \textbf{1}_{M_{UV}}(M_{{UV},i})\mathrm{PDF}_i(z,T)\mathrm{d}T,
\end{equation}

\noindent where $\textbf{1}_{M_{UV}}(M_{{UV},i})$ is an indicator function whose value is one if the argument is between $M_{UV}$ and $M_{UV}+dM_{UV}$. This probability tracks the uncertainties of the observed colours, traced by the template space, to the $z-M_{UV}$ space, including the correlation between the two variables. As an example, we show the $P_i(z,M_{UV}|I_i)$ for an arbitrary high-redshift ALHAMBRA source in Fig.~\ref{fig:PI} ({\it Top}). 

Finally, to take into account the photometric errors, and the Eddington bias effects (see Sect.~\ref{sec:photom}), we needed to convolve this matrix by the source function (Eq.~[\ref{eq:source}]). Formally:

\begin{equation}
\mathrm{PDF}_i(z,M_{UV})=P_i(z,M_{UV}|I_i)\ast S(I_0|I_i,\sigma_{I_i}).\label{eq:post}
\end{equation}

\noindent The final posterior $\mathrm{PDF}_i(z,M_{UV})$ for our example galaxy is shown in the bottom panel of Fig.~\ref{fig:PI}. We note that the PDFs show sharp edges because, to reduce the calculation time, a cut was applied to small $P_i(z,M_{UV}|I_i)$ values that were confirmed to be insignificant. To derive the limiting magnitude, $M_{{UV},lim}(z)$, corresponding to $I_0=24$, as a function of redshift (red line in Fig.~\ref{fig:PI}), we calculated at each redshift the absolute magnitude for the template that, once normalised to $I_0=24$, gives the brightest absolute magnitude at that $z$:

\begin{equation}
M_{{UV},lim}(z)=min[M_{UV}(z,T|I_0=24)].
\end{equation}

\noindent In this way we assured a 100\% complete selection above $M_{{UV},lim}(z)$. Finally, in the same Fig.~\ref{fig:PI}, we also show how the same galaxy would be seen in a traditional approach in which only the \texttt{BPZ2} redshift, $z_{\rm{ML}}$ \citep{molino14}, and the corresponding $UV$ magnitude (i.e. $M_{UV}$ corresponding to the observed $I$, $z_{\rm{ML}}$ and the best fitting template), were taken into account.

\section{Luminosity function derivation}\label{sec:lf}

In this section we derive the ALHAMBRA LF and model it in 2D, again following the scheme presented in LS17. The authors of LS17 calculate the LF separately for quiescent and star-forming galaxies. We use exactly the same code in this work. However, in the following text we skip the details related to the separation between quiescent and star-forming galaxies. This is because the integrated total number of quiescent galaxies in the ALHAMBRA catalogue at the redshift range of our interest is non-significant (0.7 to be more exact), and these galaxies can safely be ignored in our analysis. The low number of quiescent galaxies is understandable because these galaxies are too faint in their rest-frame $UV$ to be detected by ALHAMBRA with $F814W\leq 24$. Here we give a summary of the steps taken to calculate the LF. For more detailed description of the methodology, see LS17.

\subsection{2D luminosity function}

Having the $\mathrm{PDF}_i(z,M_{UV})$ for each ALHAMBRA galaxy, we are now in the position of constructing the LF simply by summing up the individual probability distributions, divided by the volumes they probe. To take into account the cosmic variance, we calculated the LF field by field. For an ALHAMBRA field $j$ the LF is given by the equation:

\begin{equation}
\Phi_j (z,M_{UV}) = \frac{1}{A_j}\sum _i \mathrm{PDF}_i(z,M_{UV})\left (\frac{\mathrm{d}V}{\mathrm{d}z}\right )^{-1} [\mathrm{Mpc}^{-3}\mathrm{mag}^{-1}],
\end{equation}

\noindent where $i$ runs over all the galaxies in the field $j$, $A_j$ is the area (in deg$^2$) of the field $j$, $\mathrm{PDF}_i(z,M_{UV})$ is given by the equation (\ref{eq:post}), and $\mathrm{d}V/\mathrm{d}z$ is the differential cosmic volume probed by one square degree.

Next, we calculated the median value of the individual fields to obtain the total ALHAMBRA $UV$ LF:

\begin{equation}
\Phi ^{\mathrm{tot}}(z,M_{UV})= \mathrm{med}\left [\Phi_j (z,M_{UV})\right ],
\label{eq:lfsum}
\end{equation}

\noindent where the index $j$ runs the $N=48$ ALHAMBRA sub-fields. We note that we calculated the median value here, instead of the mean, because we consider this measurement more robust for the log-normally distributed number counts (see Sec. \ref{sec:cos_var}). The difference between mean and median becomes important at the brightest bins, which contain only a few galaxies, in which the mean would clearly bias the LF value towards the more populated fields.

Finally, to ensure a well controlled error budget for the LF, we degraded the resolution of each $\Phi_j (z,M_{UV})$ to create a binned LF. The LF was divided in $K$ bins in the $z$-axis and $L$ bins in the $M_{UV}$ axis, where the optimal bin sizes, $\Delta z$ and $\Delta M_{UV}$, are defined in Appendix \ref{sec:binning}. Then for each bin [$z_{\mathrm{min}}=z_k - 0.5\Delta z, z_{\mathrm{max}}=z_k + 0.5\Delta z$], [$M_{UV,\mathrm{min}}=M_{UV,l} - 0.5\Delta M_{UV}, M_{UV,\mathrm{max}}=M_{UV,l} + 0.5\Delta M_{UV}$], where $k=(1,2,...K)$, $l=(1,2,...L)$, the following formula was applied:

\begin{equation}
\Phi_{j,kl} = \frac{1}{\Delta V\Delta M_{UV}}\int _{z_{\mathrm{min}}}^{z_{\mathrm{max}}}\int _{M_{UV,min}}^{M_{UV,max}}\Phi_j (z,M_{UV})\frac{\mathrm{d}V}{\mathrm{d}z}\mathrm{d}z\, \mathrm{d}M_{UV},
\label{eq:binned}
\end{equation}

\noindent where $\Delta V$ is the cosmic volume probed by one square degree at $z_{\mathrm{min}} \leq z < z_{\mathrm{max}}$:

\begin{equation}
\Delta V = \int _{z_{\mathrm{min}}}^{z_{\mathrm{max}}} \frac{\mathrm{d}V}{\mathrm{d}z}\mathrm{d}z.
\end{equation}

\noindent The binned LF is then:

\begin{equation}
\Phi_j^\mathrm{b} \equiv \Phi_j(\boldsymbol{z},\boldsymbol{M_{UV}}) = \Phi _{j,kl}\in \mathbb{R}^{K\times L},
\end{equation}

\noindent where $\boldsymbol{z}$ and $\boldsymbol{M_{UV}}$ are the vectors that define the binned histogram. Finally, the total binned LF is obtained with Eq.~(\ref{eq:lfsum}) by taking the median of the individual $\Phi_j^\mathrm{b}$.

\subsection{Luminosity function model}\label{sec:lfmodel}

We modelled the ALHAMBRA LF with a Schechter function \citep{schechter76}:

\begin{equation}
\Phi _{\mathrm{mod}}(M_{UV}|\theta _{\Phi})=0.4\mathrm{ln}(10)\phi^{\ast}\frac{10^{0.4(M_{UV}^{\ast}-M_{UV})(1+\alpha)}}{\mathrm{exp}\{10^{0.4(M_{UV}^{\ast}-M_{UV})}\}},
\label{eq:sch}
\end{equation}

\noindent where $\theta _{\Phi} = [M_{UV}^{\ast},\phi^{\ast},\alpha]$ are the parameters that define the model. $M_{UV}^{\ast}$ is the characteristic magnitude corresponding to the transition magnitude from a power law LF to an exponential LF, $\phi^{\ast}$ is the characteristic density, offering the normalisation of the LF, and $\alpha$ determines the slope of the power law variation at the faint end. 

The modelling was done in 2D. For this purpose, we calculated a model LF using Eq. (\ref{eq:sch}) at the same grid as our binned LF. The $\chi ^2$ to be minimised is then given by the equation:

\begin{equation}
\chi  ^2(\Phi |\theta _{\Phi},\Sigma _{\Phi}) = [\mathrm{ln}\Phi - \mathrm{ln}\Phi _{\mathrm{mod}}]^{\mathrm{T}}\Sigma _{\Phi}^{-1}[\mathrm{ln}\Phi - \mathrm{ln}\Phi _{\mathrm{mod}}],
\label{eq:chi2}
\end{equation}

\noindent where $\Sigma _{\Phi}$ is the LF covariance matrix defined in Sect. \ref{sec:errors} below, and the $\chi ^2$ is defined in log-space because the LF values follow a log-normal distribution rather than a Gaussian one (see Sect. \ref{sec:bias}). The posterior distribution of the model parameters is then:

\begin{equation}
P(\theta _{\Phi}|\Phi, \Sigma _{\Phi}) \propto \mathrm{exp}(-\frac{1}{2}\chi ^2)P(\theta _{\Phi}),
\label{eq:theta_prob}
\end{equation}

\begin{figure}
\centering
\includegraphics{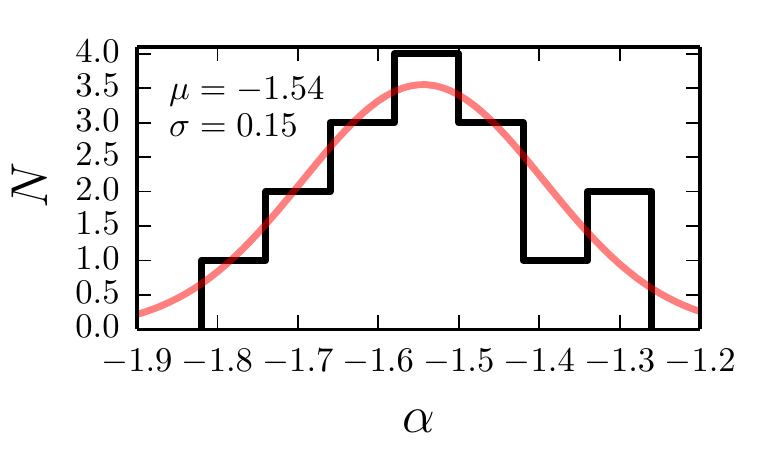}
\caption{Distribution of $\alpha$ values as compiled from the literature by \citet{parsa16} at redshifts $z=2.7-4.0$. A Gaussian fit to this distribution is shown as a red line and the median and sigma of this Gaussian are labelled in the panel.} 
\label{fig:alphas}
\end{figure}

\noindent where the distribution is normalised to unity and $P(\theta _{\Phi})$ is the prior in the parameters. Here we have assumed an uninformative, flat prior, $P(\theta _{\Phi}) = 1$, for $M_{UV}^{\ast}$ and $\phi^{\ast}$. We do not have enough data points at the faint end to accurately trace $\alpha$. Therefore, for $\alpha$ we needed to rely on a prior information. \citet{parsa16} offers a compilation of different $\alpha$ values found in the literature. From their table we selected the $\alpha$ values at the redshift range of our interest, ending up with 16 values in the range $z=2.7-4.0$, and created a histogram of them. We fitted the histogram with a Gaussian function, resulting in median and standard deviation of $\alpha = -1.54\pm 0.15$, see Fig. \ref{fig:alphas}. Finally, we used this as a prior distribution for the $\alpha$ parameter.

\subsection{Final modelling}\label{sec:lfmod}

The 2D LF modelling was then carried out by minimising the equation (\ref{eq:chi2}). This was done using the \texttt{emcee} code which provides a collection of solutions in the parameter space, with the density of the solutions being proportional to the posterior probability of the parameters, meaning that it empirically maps Eq. (\ref{eq:theta_prob}). The most probable values of the parameters and their uncertainties were then obtained as the median and the dispersion of the projected solutions.

\section{Luminosity function uncertainty estimates}\label{sec:errors}

The LF uncertainty has two dominant terms: the statistical  (i.e. the shot noise term), and the cosmic variance term \citep{robertson10,smith12}. Because of the uncertainties in both the redshifts and the magnitudes, the values of the LF in different bins are correlated in both dimensions. Following again LS17, we estimated both the shot noise and cosmic variance terms of the covariance matrix for the LF $\Phi$. The covariance matrix is given in relative form, as the LF is finally fitted in logarithmic space.

\subsection{Shot noise term}

The shot noise term was derived with the bootstrapping technique \citep{davison97}. For this purpose, 20000 bootstrap samples of the median LF were created, noted as $\Phi_p^{\mathrm{tot}}, p=(1,2,...20000)$. The shot noise term of the covariance matrix $\Sigma$ is then given by the equation:

\begin{multline}
\Sigma _S \equiv \Sigma _S(z_m,z_n,M_{{UV},q},M_{{UV},l})=\\ \frac{\mathbb{E}[\Phi _{p}^{\mathrm{tot}}(z_m,M_{{UV},q})\Phi _{p}^{\mathrm{tot}}(z_n,M_{{UV},l})]}{\Phi _{p}^{\mathrm{tot}}(z_m,M_{{UV},q})\Phi _{p}^{\mathrm{tot}}(z_n,M_{{UV},l})}-1,
\end{multline}

\noindent where $\mathbb{E}$ is the expected value (i.e. the mean) operator, the indices $m$ and $n$ run the redshift bins, and the indices $q$ and $l$ run the absolute magnitude bins. The covariance between luminosity bins at a given redshift is mapped by setting $m=n$, and the covariance between redshift bins at the given magnitude by setting $q=l$.

\subsubsection{Cosmic variance term}\label{sec:cos_var}

The large scale density fluctuations of the Universe lead to field to field variations in the observed galaxy number counts. This cosmic variance often causes uncertainties larger than the shot noise term derived in the previous section. Here we have derived the cosmic variance term of the LF covariance matrix following the prescription given in LS17 and \citet{lopez-sanjuan15}.

The relative cosmic variance is defined as \citep{somerville04}:

\begin{equation}
\sigma_v = \frac{\langle n^2 \rangle - \langle n \rangle^2}{ \langle n \rangle^2} - \frac{1}{ \langle n \rangle},
\end{equation}

\noindent where $\langle n \rangle$ and $\langle n^2 \rangle - \langle n \rangle^2$ are the mean and the variance, respectively, of the number density distribution of galaxies in the 48 ALHAMBRA subfields. The number density in each ALHAMBRA field and for each redshift-magnitude bin is obtained by equation (\ref{eq:binned}). This is then fitted by a log-normal function, whose dispersion encodes both the dispersion due to the Poisson shot noise and the intrinsic dispersion due to the galaxy clustering. The latter, (i.e. the cosmic variance term), is separated from the former using a maximum likelihood estimator \citep[MLE, see][]{lopez-sanjuan15,lopez-sanjuan15b}. MLE offers both the value of the cosmic variance term and its error.

Ideally, one would derive the cosmic variance at exactly the same bins as the LF. However, due to the limited amount of galaxies in our sample, we have calculated the variance in two magnitude bins at $z=2.5-3.5$, and one magnitude bin at $z=3.5-4.5$, and derive from modelling the cosmic variance, $\sigma_{v,mod}$, at the same resolution as our LF (see Sect. \ref{sec:bias}). The cosmic variance term of the covariance matrix is then given by the equation:

\begin{multline}
\Sigma _v \equiv \Sigma _v(z_m,z_n,M_{{UV},q},M_{{UV},l}) = \\\delta _{mn} \frac{\sigma _{v,mod}(z_m,M_{{UV},q})\sigma _{v,mod}(z_n,M_{{UV},l})]}{\sqrt{V _{\mathrm{eff}}(z_m,M_{{UV},q})V _{\mathrm{eff}}(z_n,M_{{UV},l})}}\sqrt{\Delta V_m\Delta V_n},
\label{eq:sigma_v}
\end{multline}

\noindent where the Kronecker $\delta _{mn}$ is one if $m=n$ and zero otherwise, implying that the redshift bins are independent, and the effective volume, $V_{\mathrm{eff}}$, takes into account the smaller cosmic volume probed by the bins affected by our selected magnitude limit $I_0=24$:

\begin{multline}
V_{\mathrm{eff}}(\boldsymbol{z},\boldsymbol{M_{UV}})=\\ \frac{1}{\Delta M_{{UV},q}}\int _{z_{m,\mathrm{min}}}^{z_{m,\mathrm{max}}}\int _{M_{UV,q,min}}^{\mathrm{min}[M_{UV,q,max},M_{UV,lim}]}\frac{\mathrm{d}V}{\mathrm{d}z}\mathrm{d}z\,\mathrm{d}M_{UV}.
\end{multline} 

\subsection{Final covariance matrix}
The final total covariance matrix was obtained by simply summing the shot noise and the cosmic variance terms:

\begin{equation}
\Sigma _{\Phi} = \Sigma _S + \Sigma _v\,.
\end{equation}

\noindent This covariance matrix provides the LF error estimate, mapping the correlations due to the redshift and magnitude uncertainties.

\begin{figure}
\centering
\includegraphics{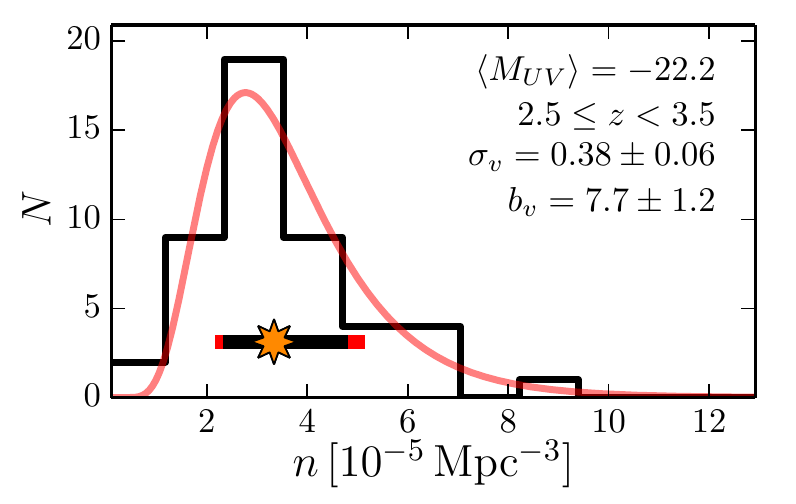}
\includegraphics{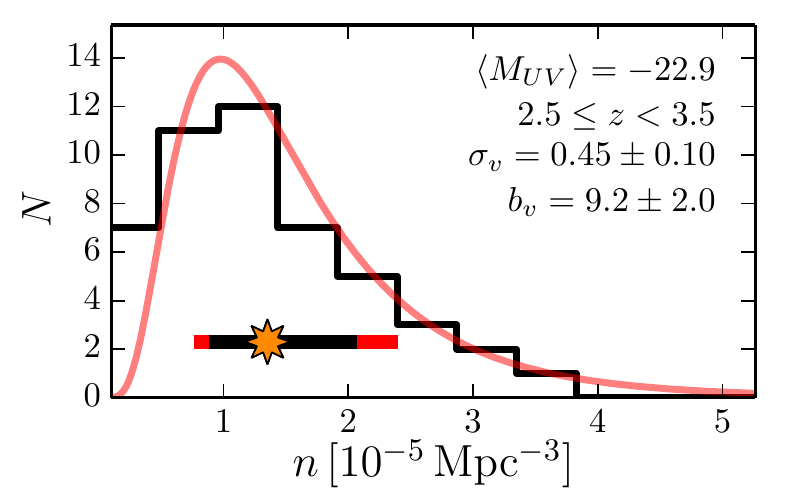}
\includegraphics{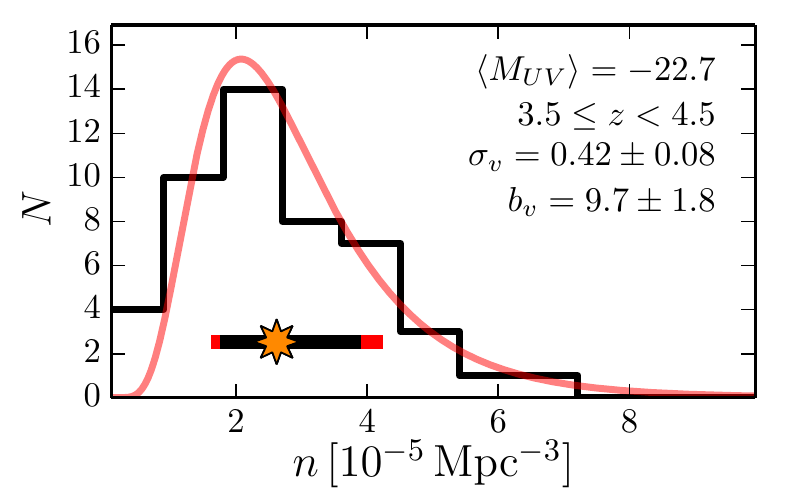}
\caption{Number density distribution in the 48 ALHAMBRA sub-fields in three magnitude-redshift bins. The mean magnitude and the redshift range of each bin are given in the panels. The total dispersion is shown as a red bar while the star and the black bar mark the median and the intrinsic dispersion retrieved by the MLE, respectively. The red solid lines show the best MLE solutions convolved with the Poisonian errors and are independent of the histogram binning. The derived values of the relative cosmic variance and the corresponding bias values are labelled in the panels.}
\label{fig:cosvar}
\end{figure}

\section{Galaxy bias}\label{sec:bias}

The galaxy bias, in addition to being an interesting result by itself, giving information about the clustering of the galaxies, is necessary for our LF modelling as it enters the LF covariance matrix calculation through its cosmic variance term. The galaxy linear bias can be derived from the cosmic variance (see Sect. \ref{sec:cos_var} above) by the equation \citep{moster11}:

\begin{equation}
b_v (\boldsymbol{z},\boldsymbol{M_{UV}}) =\frac{\sigma _v(\boldsymbol{z},\boldsymbol{M_{UV}})}{\sigma _{v,dm}(\boldsymbol{z})},
\label{eq:bias}
\end{equation}

\noindent where $\sigma _{v,dm}$ is the cosmic variance of the dark matter calculated at the redshift bins of our interest, for an area of 0.051 deg$^2$, the median area of the 48 ALHAMBRA sub-fields. This equation is based on an assumption that the bias does not depend on the scale of structure. This theoretical cosmic variance was computed in each volume using the code QUICKCV\footnote{QUICKCV is available at www.phyast.pitt.edu/$\sim$janewman/quickcv} based on work published in \citet{newman02}. The code computes the cosmic variance from the dark-matter power spectrum using a window function which is one inside the volume of interest and zero otherwise. The dark-matter power spectrum at each redshift bin was obtained using the CAMB software \citep{lewis00}, including the non-linear corrections of HALOFIT \citep{smith03}.

In Appendix \ref{sec:binning} we introduce the optimal binning for the bias calculation, ending up with three redshift-magnitude bins. The number density distributions of galaxies in the 48 ALHAMBRA subfields, the best MLE solutions, and the derived cosmic variances and biases together with their uncertainties in these three bins are shown in Fig. \ref{fig:cosvar}. We see that the bias values vary in the range $b_v\sim 8-10$. These bias values are large, actually larger than any previous measurement at these redshifts. However, considering the high redshift and brightness of our galaxies these bias values are reasonable, as discussed in the following.

\begin{figure*}
\centering
\includegraphics{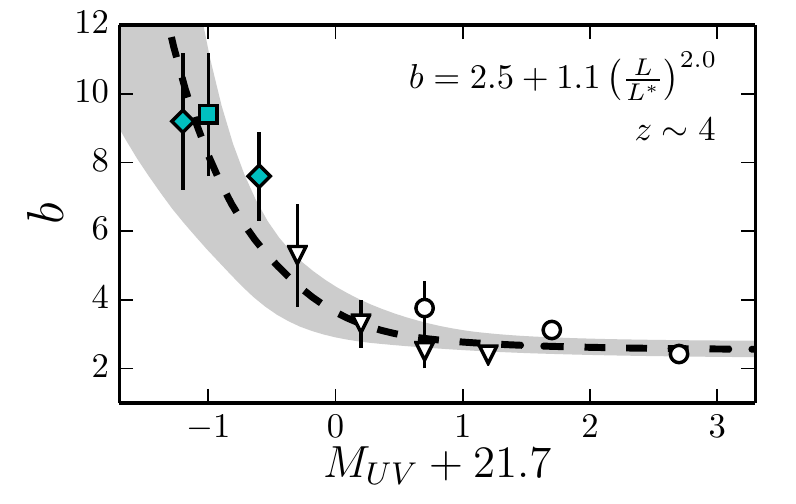}\includegraphics{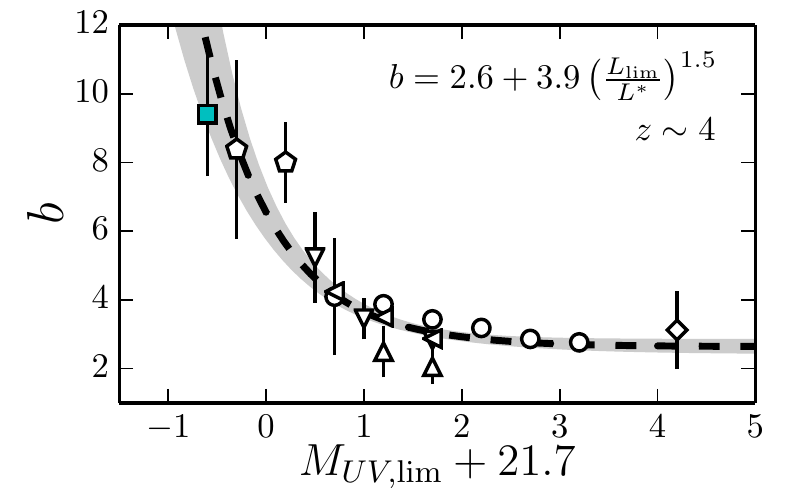}
\caption{ {\it Left:} Bias values as a function of absolute $UV$ magnitude. The blue filled square represents our data point at $z\sim 4$ and blue filled diamonds our data points at $z\sim 3$. These are shown as a comparison but are not used in the fit. The data points from \citet{ouchi04a} and \citet{cooray06} are shown as triangles and circles, respectively. {\it Right:} Bias values as a function of the limiting absolute $UV$ magnitude of each data set. The blue filled square shows our data point at $z\sim 4$. The results from the literature are shown as circles \citep{ouchi05}, pentagons \citep{allen05}, triangles \citep{ouchi01}, inverted triangles \citep{ouchi04}, diamonds \citep{arnouts02}, left-pointing triangles \citep{hildebrandt09}, stars \citep{harikane16}, and hexagons \citep{harikane17}. The black dashed line shows our best fit of Eq. (\ref{eq:biasmod}) as also labelled in the figures. The 1-$\sigma$ errors are shown as grey shaded areas.} 
\label{fig:biasfit}
\end{figure*}

\subsection{Bias function model}\label{sec:bmodel}

At the redshift range of our interest, the galaxy bias is found to increase both with redshift and with luminosity \citep[e.g.][]{hildebrandt09}. Our data allows us to robustly estimate the bias in only three magnitude-redshift bins, while the magnitude dependence of bias at $z\sim 3$ is, to our knowledge, poorly studied in the literature. We opted for tracing the galaxy bias as a function of magnitude at $z\sim 4$, where previous studies provide data points at fainter magnitudes, and applying this dependence for both our redshift bins. We combined our data with the literature data at $z\sim 4$ where the bias information is given for magnitude bins (see Fig. \ref{fig:biasfit}, {\it Left}) and fit a relation:

\begin{equation}
b_{mod} (M_{UV})= A + B\left ( \frac{L}{L^{\ast}}\right )^C,
\label{eq:biasmod}
\end{equation}

\noindent where $L^{\ast}$ is the luminosity corresponding to a normalisation magnitude $M^{\ast}=-21.7$, and $A$, $B$, and $C$ are the fitted constants, $A=2.5\pm 0.2$, $B=1.1\pm 0.8$, and $C=2.0\pm 0.8$. To show that the $z\sim 3$ bias values are also well traced by this model, we have plotted them in Fig. \ref{fig:biasfit}, {\it Left}. However, these points are not used in the fit. Most of the literature data is given as a function of the limiting magnitude of the corresponding survey, instead of the magnitude bin. For comparison, we also show this plot with our $z\sim 4$ data points included in Fig. \ref{fig:biasfit}, {\it Right}. The constants for this fit are $A=2.7\pm 0.2$, $B=3.2\pm 0.2$, and $C=0.9\pm 0.1$.

We see in both panels of Fig. \ref{fig:biasfit} that the bias values we measure are higher than the values found in the literature. Considering the fact that our data traces brighter magnitudes than any of the previous studies, and considering the increasing trend of the bias with luminosity, the high values we measure are not surprising. We further study the reliability of the measured bias values in Sect. \ref{sec:halom}.

\begin{figure}
\centering
\includegraphics{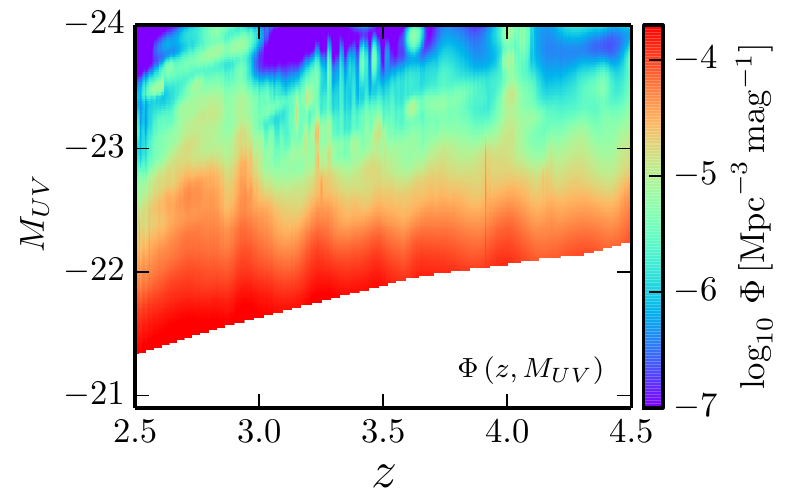}
\caption{Total differential LF of ALHAMBRA galaxies at $z=2.5-4.5$.}
\label{fig:LF2d}
\end{figure}

\begin{figure*}
\centering
\includegraphics{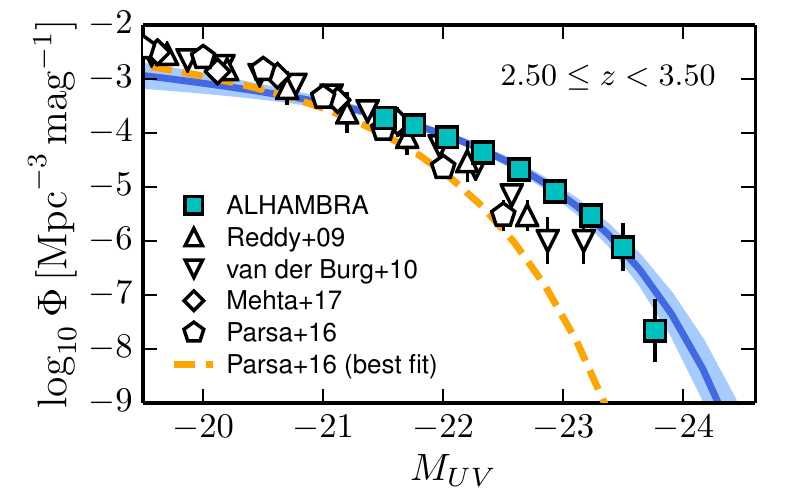}\includegraphics{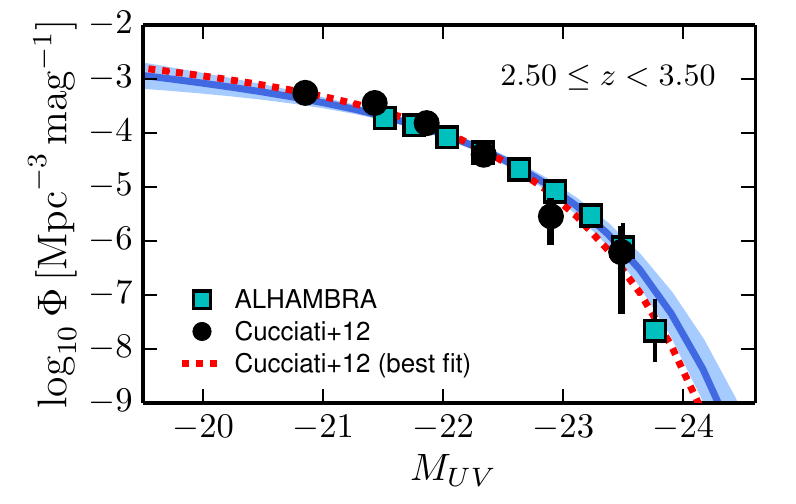}
\includegraphics{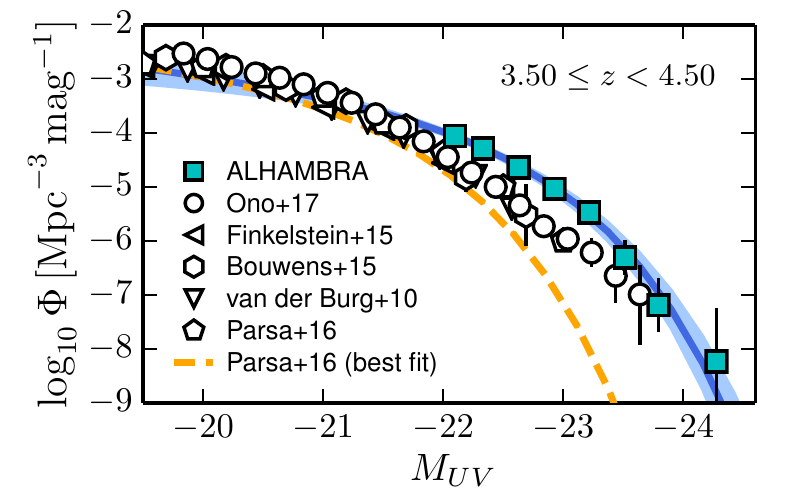}\includegraphics{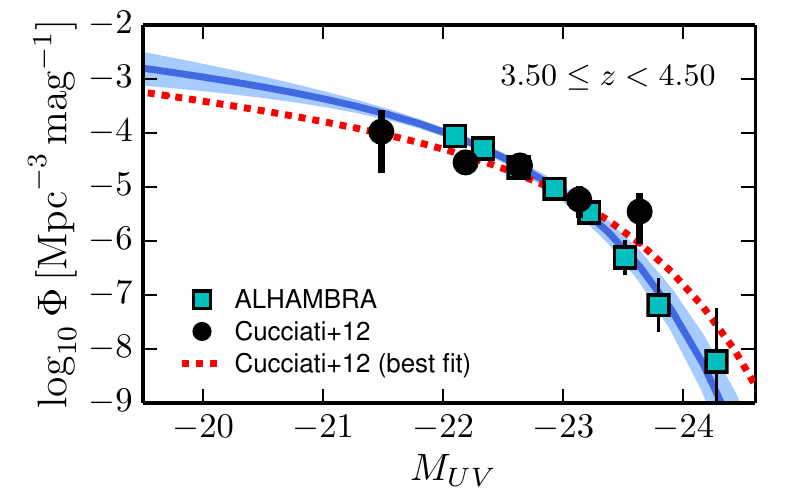}
\caption{ALHAMBRA $UV$ LF at two redshift bins as indicated in the panels. The ALHAMBRA measurements are shown as blue squares together with their 2-$\sigma$ error bars, and the blue line shows our best median Schechter fit, shaded area enclosing 95\% of the solutions. {\it Left panels:} Comparison with literature data from photometrically selected samples of high-$z$ galaxies is shown as follows: triangles \citep{reddy09}, upside-down triangles \citep{vandenburg10}, diamonds \citep{mehta17}, pentagons and orange dashed-line \citep{parsa16}, open circles \citep{ono17}, left pointing triangles \citep{finkelstein15}, and hexagons \citep{bouwens15}. {\it Right panels:} Comparison with the spectroscopic LF estimate of \citet{cucciati12} (black dots and red dotted line).}  
\label{fig:LFs}
\end{figure*}

\section{Results: ALHAMBRA $UV$ luminosity function}\label{sec:results}

In this section we present the final 2D and discretised $z\sim 3$ and $z\sim 4$ ALHAMBRA $UV$ LFs and compare them with the LFs from the literature, Sect. \ref{sec:lumfun}. We discarded a strong influence of low-redshift galaxies (Sect. \ref{sec:lowz}), and quasars (QSOs, Sect. \ref{sec:qsos}) on our LFs and compare them with the recent LF estimates in radio, Sect. \ref{sec:radio}. Finally, we derived the FUV comoving luminosity density in $z\sim 3$ and $z\sim 4$, Sect. \ref{sec:ld}.

\begin{table*}

\caption{ALHAMBRA $UV$ LF $\Phi\,(z,M_{UV})$.}

\label{lfuv_tab}

\begin{center}

\begin{tabular}{@{\extracolsep{2pt}}cccccc@{}}
\hline\hline\noalign{\smallskip}
$M_{UV}^{-}$ & $M_{UV}^{+}$ & \multicolumn{2}{c}{$2.5 \leq z < 3.5$} & \multicolumn{2}{c}{$3.5 \leq z < 4.5$}\\\noalign{\smallskip}\cline{3-4}\cline{5-6}\noalign{\smallskip}
& & $\langle M_{UV} \rangle$ & $\log_{10}\Phi$ & $\langle M_{UV} \rangle$ & $\log_{10}\Phi$\\
\noalign{\smallskip}
\hline
\noalign{\smallskip}
$-26.0$ & $-24.0$ & $-24.05$ & $-11.09 \pm 0.63$ & $-24.28$ & $-8.24 \pm 0.49$\\ $-24.0$ & $-23.7$ & $-23.76$ & $-7.66 \pm 0.29$ & $-23.79$ & $-7.19 \pm 0.25$\\ $-23.7$ & $-23.4$ & $-23.50$ & $-6.12 \pm 0.22$ & $-23.51$ & $-6.31 \pm 0.17$\\ $-23.4$ & $-23.1$ & $-23.23$ & $-5.53 \pm 0.11$ & $-23.22$ & $-5.47 \pm 0.09$\\ $-23.1$ & $-22.8$ & $-22.93$ & $-5.08 \pm 0.07$ & $-22.93$ & $-5.03 \pm 0.06$\\ $-22.8$ & $-22.5$ & $-22.63$ & $-4.67 \pm 0.04$ & $-22.63$ & $-4.64 \pm 0.04$\\ $-22.5$ & $-22.2$ & $-22.33$ & $-4.36 \pm 0.03$ & $-22.33$ & $-4.28 \pm 0.03$\\ $-22.2$ & $-21.9$ & $-22.03$ & $-4.08 \pm 0.02$ & $-22.10$ & $-4.06 \pm 0.03$\\ $-21.9$ & $-21.6$ & $-21.76$ & $-3.85 \pm 0.02$ & $\cdots$ & $\cdots$\\ $-21.6$ & $-21.3$ & $-21.51$ & $-3.72 \pm 0.03$ & $\cdots$ & $\cdots$\\
\hline
\end{tabular}
\end{center}
\tablefoot{The units of the LF are Mpc$^{-3}$\,mag$^{-1}$. The quoted uncertainties only reflect the diagonal terms of the covariance matrix $\boldsymbol{\Sigma}_{\Phi}$, both shot noise and cosmic variance.}
\label{tab:lf}
\end{table*}

\begin{table*}
\caption[]{Best fitting Schechter parameters, their correlation coefficients, and the luminosity density derived from the LFs at our two redshift bins.}
\label{tab:ourvals}
\centering
\begin{tabular}{c c c c c c c p{15mm}}
\hline
$z$ & $M_{UV}^\ast$ & \begin{tabular}{@{}c@{}}$\phi ^\ast$\\ $[$/10$^3$ Mpc$^3]$\end{tabular} & $\alpha$ & $\rho _{M_{UV}^\ast \phi}$ & $\rho _{M_{UV}^\ast \alpha}$ & $\rho _{\phi \alpha}$ & \begin{tabular}{@{}c@{}}log(LD)\\ $[$W/Hz/Mpc$^3]$\end{tabular}\\
\hline
2.5-3.5 & $-21.62\pm 0.11$ & $0.51\pm 0.07$ & $-1.53\pm 0.15$ & 0.00065 & 0.92 & 0.84 & 19.28$^{+0.15}_{-0.11}$\\
3.5-4.5 & $-21.67\pm 0.10$ & $0.54\pm 0.08$ & $-1.65\pm 0.17$ & 0.00079 & 0.80 & 0.50 & 19.44$^{+0.26}_{-0.18}$\\
\hline
\end{tabular}
\end{table*}

\subsection{The luminosity function}\label{sec:lumfun}

The 2D ALHAMBRA LF for the whole redshift range of our interest, $z=2.5-4.5$, consisting of an integrated total number of galaxies of 3861.5, is shown in Fig.~\ref{fig:LF2d}. One can observe over-dense strips in redshift space, reflecting the presence of cosmic structures. Also the $z - M_{UV}$ correlations (Fig. \ref{fig:PI}) are visible especially at the brighter magnitudes where the density of objects is lower. The incompleteness at the faint end due to volume effect is evident, but is properly taken into account in the modelling.

The optimal binning of this LF is derived in Appendix \ref{sec:binning}. We find that $\Delta z=1$ and $\Delta M_{UV}=0.3$ provide a proper error budget, both shot noise and cosmic variance. We show in Table \ref{tab:lf} the values of the resultant LF in the two redshift bins, $z=2.5-3.5$, and $z=3.5-4.5$. For the magnitude bins, we show both the bin edges and the number weighted median values. The best fitting Schechter parameters are given in Table \ref{tab:ourvals}. We also show the correlation coefficients, $\rho$, between each pair of values $M_{UV}^\ast$, $\phi ^\ast$, and $\alpha$. We see that while $M_{UV}^\ast$ and $\phi ^\ast$ are quite independent of each other, both $M_{UV}^\ast$ and $\phi ^\ast$ are highly correlated with $\alpha$. We plot these LFs, together with the best fitting model, in Fig. \ref{fig:LFs}. For comparison, we also show LF data given in the literature. One one hand, we compare our LFs with the literature data originating from works in which the galaxy selection is based on broad-band photometry, either applying drop-out selection or a selection made in photometric redshifts (left panels). On the other hand, we overplot the literature data derived from the spectroscopic VVDS survey \citep[][right panels]{cucciati12}.

In the left panels, we see a discrepancy between our LFs and the LFs from the literature, our data showing an excess of objects at the bright end. However, in the right panels we see that our data agree very well with the spectroscopic results. Hence, even though being discrepant at its bright end with most of the literature data, our results are compatible with those of the VVDS, to our knowledge the only magnitude-limited spectroscopic study at these redshifts.

It has already been discussed in the literature \citep[e.g.][]{paltani07,lefevre15,inami17} that the colour-colour selection leaves out a fraction of genuine high-$z$ galaxies which might explain the disagreement of our results with these studies. On the other hand, a selection made in photometric redshifts can suffer from similar biases, the quality of the photo-zs depending strongly on, for example, the number of bands used for their derivation \citep{benitez09}. We showed in \citet{viironen15} that even in the case of ALHAMBRA with 23 bands, a strictly photo-z selected sample of galaxies lies within a typical LBG selection box and hence also misses the kind of galaxies shown to exist outside the box. In the same article, we also showed that when the information in the whole redshift PDFs is taken into account, an important fraction of galaxies is found outside the colour-colour selection boxes.
 
At the redshift bin $3.5 < z < 4.5$, we observe that the volume probed by ALHAMBRA cannot explain the excess of bright galaxies, because the recent study by the GOLDRUSH project \citep{ono17} covers a much larger area, $\sim 100$ deg$^2$. On the other hand, our brightest bins with $M_{UV}\lesssim -23.5$ closely agree with those of \citet{ono17}. Hence, the main difference between their results and ours is that our data points to a Schechter-shaped LF with a brighter $M_{UV}^{\ast}$ by $\sim 0.75$ magnitudes, while their data favours a double power-law. We also see in the left panels of Fig. \ref{fig:LFs} that at the faint end our data points approach those from the literature. This implies an excess of galaxies only at magnitudes close to the $M^{\ast}_{UV}$. We study in the following the possible causes of the observed excess. We concentrate here on the redshift range $z\sim 4$ where the recent \citet{ono17} results exist.

\begin{figure}
\centering
\includegraphics{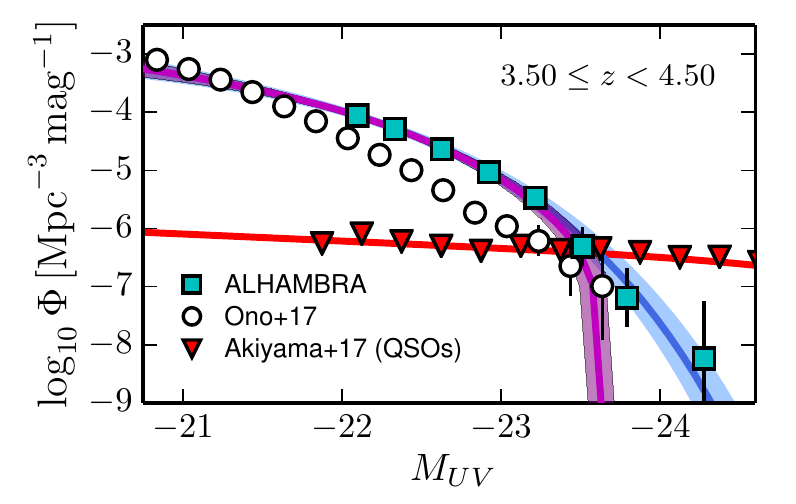}
\caption{ALHAMBRA $UV$ LF of Fig. \ref{fig:LFs} at redshift $z\sim 4$ and the corresponding Schechter fit (blue squares and blue line). The LF from \citet{ono17} is shown as white circles. The QSO LF of \citet{akiyama17} at $z\sim 4$ is plotted as red upside-down triangles. The purple line shows the best Schechter fit when this (absolute maximum) QSO contamination is subtracted from our $z\sim 4$ LF fit.}  
\label{fig:LF_extra}
\end{figure}

\subsection{Bright-end excess; low redshift galaxies}\label{sec:lowz}

We observe in Fig. \ref{fig:LFs} an excess of objects close to $M^{\ast}_{UV}$ as compared to studies in which either drop-out or photo-z selected samples derived from broad-band data are used. Our results agree with the spectroscopic VVDS LF reinforcing our confidence on the ALHAMBRA redshift PDFs and on the methodology used in this paper.

However, a natural concern is if low redshift galaxies could cause this excess. Some galaxies at low redshift can be confused with the high-redshift ones because of the Lyman/Lyman-$\alpha$ break vs. 4000 \AA~break degeneracy, causing, in simplified terms, double peaked redshift PDFs. The low-z galaxies are more numerous than the high-z ones, hence, one can worry if, in our approach of summing up the redshift PDFs, these double-peaked PDFs could cause a net flux of low-z objects towards high-z and explain the observed excess in our LF. We studied how the number of galaxies at $z\sim 4$ would change when we simply account for all the galaxies with the maximum likelihood photo-z, that is, 'the primary peak', in the range $3.5 \leq z \leq 4.5$ and compare this to the same number derived from the LF calculated in this paper. The number we find is 1287 maximum likelihood photo-z objects vs. 1138.8 objects when the information from the whole redshift PDF is taken into account. In other words, our approach of integrating the PDFs, and hence taking into account also the 'secondary peak objects', or objects even with only very little probability to be at the redshift of interest, leads to a smaller number of high-z objects than the approach of selecting the 'primary peak objects' and summing them up does. Hence, it seems clear that the secondary peak objects do not cause a net flux of low-z objects towards high-z.

To further quantify the different components influencing our LF at $z\sim 4$, we calculate the contribution from objects i) with primary peak - the peak corresponding to the maximum value of the redshift PDF - at $3.5 \leq z \leq 4.5$, ii) with primary peak at $z<2$, but secondary peak at $3.5 \leq z \leq 4.5$, and iii) with primary peak at $z>2$ and some probability (but not the peak) at $3.5 \leq z \leq 4.5$. The contributions from the cases i), ii), and iii) are 60\%, 30\%, and 10\%, respectively. Hence, our LF at $z\sim 4$ is dominated by objects with primary peak at the corresponding redshift interval, with an additional contribution of 30\% caused by objects with primary peak at low-z. We recall here again, that it has been shown that genuine high-z objects may come out with redshift PDFs with primary peak at low-z \citep{paltani07,lefevre15}. Finally, a small contribution of 10\% from objects with primary peak at $z>2$, but outside the $z\sim 4$ redshift interval is also accounted for. This exercise shows us that our LF is not simply caused by accumulation of secondary peaks of (more numerous) low-z galaxies, but is indeed dominated by objects with primary redshift peak at the range of interest and an additional contribution from the less certain high-z objects (i.e. those with primary peaks outside the range of interest). \citet{lefevre15} discusses, in the framework of the spectroscopic VUDS (The Vimos Ultra-Deep) Survey that in their sample 17.5\% of genuine high-z galaxies at $z>2$ have a primary peak at low redshift. They also state that the actual fraction shows a variation with redshift. In addition to the redshift variation, the photometric redshift code used in their analysis is different from ours, so that an exact match in the percentages should not be expected.

Finally, we note that the large bias values we derived for our galaxies in Sect. \ref{sec:bias} are not compatible with the observed biases at low-z \citep[see e.g.][]{arnalte14} and further support the genuine high-$z$ nature of these galaxies.

\subsection{Bright-end excess; quasar contamination}\label{sec:qsos}

Of the above literature LF studies \citet{reddy09} and \citet{ono17} discuss the importance of QSO contamination of drop-out samples at the bright end of the LF and correct for it, while for the VVDS spectroscopic study \citep{cucciati12} this is not an issue (only spectra with galaxy quality flags are used). As  discussed in \citet{viironen15}, we do not expect a significant contamination of QSOs in our data. This is because we removed the stellar-like objects from our sample and the BPZ templates do not include QSOs, and consequently the QSO redshifts are poorly defined. However, if there were any QSOs contaminating our LF, this would occur mainly at the bright end where, on one hand, galaxy number counts are lower, and on the other hand, the faintest and thus the most numerous QSOs reside. For these reasons we further study here if these objects could be causing the bright-end excess.

Recently a catalogue of ALHAMBRA type I AGN reaching m(F814W) = 23 was presented in \citet{chaves-montero17}. We ran our photometric pre-selection and LF calculation code on this catalogue and corrected the numbers found by the completeness values given in \citet{chaves-montero17}. As the QSO redshifts are poorly defined, we might also have contamination by QSOs at other redshifts than those which are our focus here. It was shown in \citet{viironen15} that a small contamination from low-z QSOs can be expected. To be conservative, we used the lowest completeness value given by \citet{chaves-montero17}, 67 \%, given for the $z < 2$ QSOs with three detected emission lines. For QSOs at higher redshift and with only two detected lines the completeness is actually better. We find that some of the QSOs indeed enter our selection and contaminate the bright end. However, this contamination is always below $\sim 20\%$, which corresponds to $< 0.1$ dex in our LF, hence being smaller than the symbol size in Fig. \ref{fig:LFs}. Thus, the QSO contamination of our LFs is negligible and cannot explain the bright-end excess.

To further support this claim, we show in Fig. \ref{fig:LF_extra} our LF at $z\sim 4$, together with the corresponding LF from \citet{ono17} and the QSO LF from \citet{akiyama17}. We also show the Schechter fit to our data and the Schechter resulting from subtracting the QSO LF from our fit. We note that this would be an absolute upper limit for the QSO contamination because only a fraction of the existing QSOs enter our sample as discussed above and in \citet{viironen15}. However, even this absolute maximum contamination could not explain the observed excess around $M^{\ast}_{UV}$.

\begin{figure}
\centering
\includegraphics{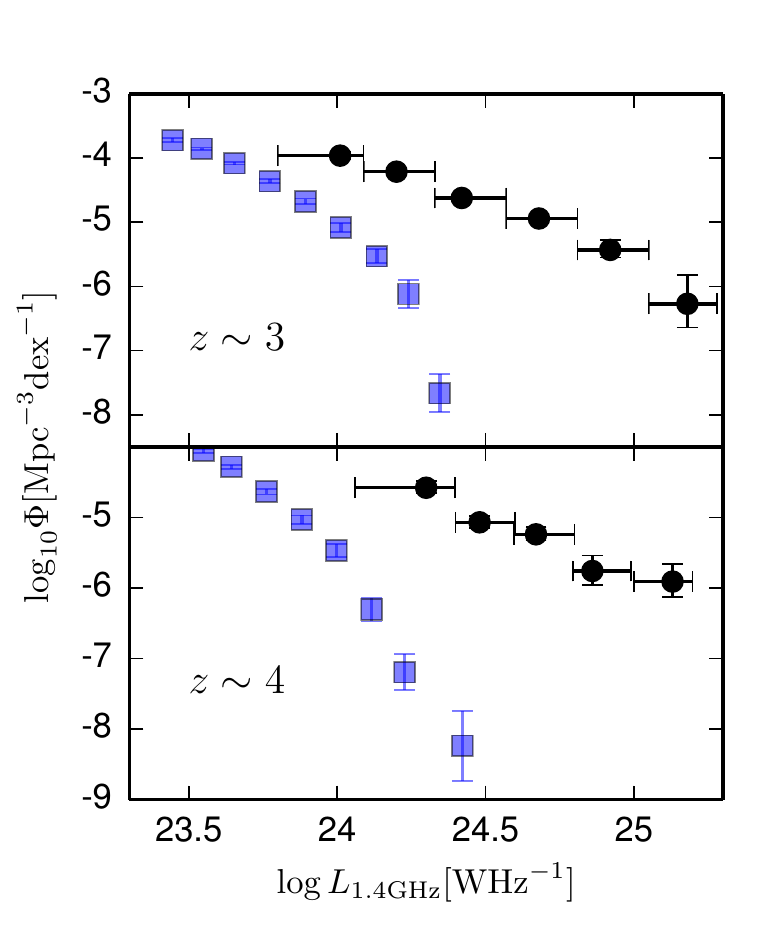}
\caption{Radio LF. The radio measurements of \citet{novak17} are shown as black dots. Our $UV$ luminosities converted to radio following the prescription of \citet{novak17} are shown as blue squares.}
\label{fig:rad}
\end{figure}

\subsection{Comparison with radio data}\label{sec:radio}

We find above that our LFs predict an excess of bright objects as compared to most of the literature data, being however compatible with spectroscopic data. We also find that low-z galaxies nor QSOs cannot explain this excess. On the other hand, the $UV$ LFs do not give a complete picture of the high redshift star-forming galaxies because part of the galaxies are probably missed, and all are faded, due to the presence of dust \citep[see][and references therein]{mancuso16}. In this section we compare our LF, once corrected for extinction, with the recent radio LF estimate of high-mass and highly star-forming galaxies \citep{novak17}. The non-thermal radio emission offers a dust-unbiased view of the star formation function and should offer the upper limit of the LF at its bright end.

To compare our $z\sim3$ and $z\sim4$ LFs with the \citet{novak17} radio LFs, we converted our $UV$ luminosities to radio luminosities using the expression given by them:

\begin{multline}
\log\left (\frac{L_{1.4\mathrm{GHz}}}{\mathrm{WHz}^{-1}}\right )=\\16.556-0.4(M_{{UV},1600}-A_{UV})-q_{\mathrm{TIR}}(z),
\end{multline}

\noindent where $L_{1.4GHz}$ is the radio luminosity, $q_{\mathrm{TIR}}=2.78\times (1+z)^{-0.14}$ links the radio luminosity to the total infrared luminosity \citep[see][and references therein]{novak17}, and $A_{UV}$ is the extinction at ultraviolet wavelength. As our luminosity distribution of galaxies is very similar to that of \citet{cucciati12}, we use the extinction values given by them: for $z=2.5-3.5$, $A_{UV}=1.47$, and for $z=3.5-4.5$, $A_{UV}=0.97$. We also added a small correction, $\Delta M_{UV}=+0.035$ to our luminosity values in order to correct them from 1500 \AA~to 1600 \AA. This was done by roughly defining the average $\beta$-slopes for our galaxies, giving $\beta \sim -1.7$, and deriving the correction from there. The comparisons of the resultant radio LFs are shown in Fig.~\ref{fig:rad}.

We see that our bright end prediction is still well below the radio based LF estimate at the bright end. This is not surprising, because the $UV$ surveys under-sample the highest SFR galaxies \citep[see][and references therein]{mancuso16}. So, despite the fact that we observed more bright galaxies than most of the studies in the literature, comparison with radio data reveals that we still are missing a significant number of very bright galaxies obscured by dust.

\begin{figure*}
\centering
\includegraphics[width=14cm]{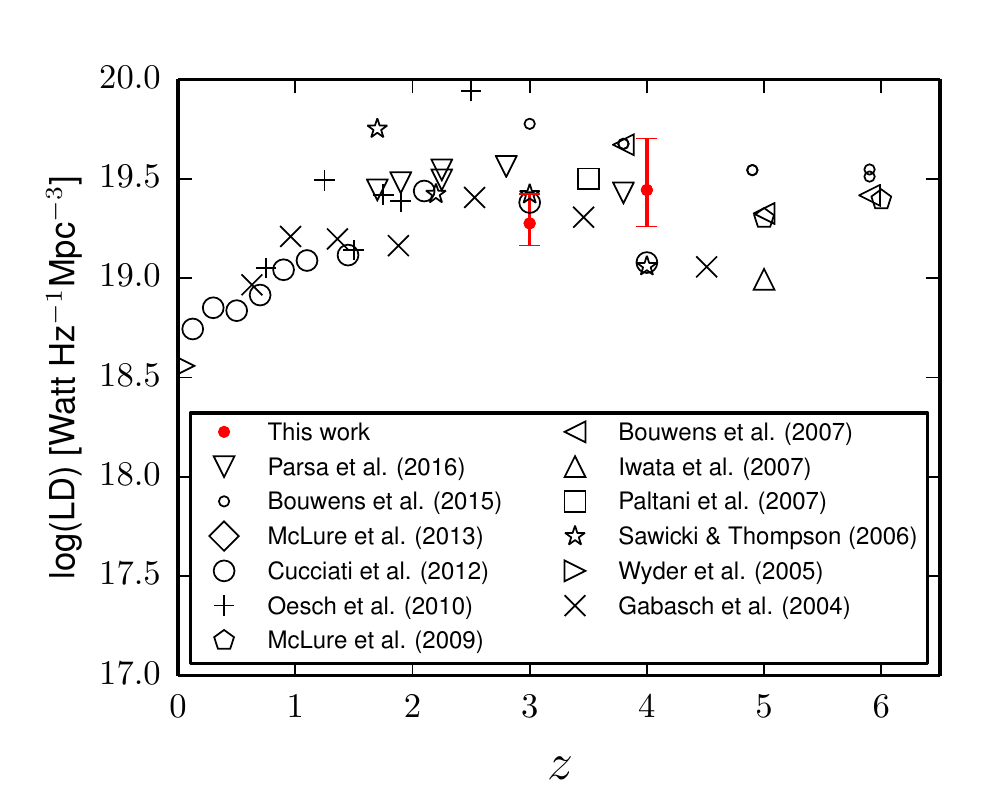}
\caption{Luminosity density derived integrating the Schechter functions given by us and given in the literature: \citet{parsa16}, \citet{bouwens15}, \citet{mclure13}, \citet{cucciati12}, \citet{oesch10}, \citet{mclure09}, \citet{bouwens07}, \citet{iwata07}, \citet{paltani07}, \citet{sawicki06}, \citet{wyder05}, and \citet{gabasch04}.}
\label{fig:ld}
\end{figure*}

\subsection{The FUV comoving luminosity density}\label{sec:ld}

Generally the Schechter parameters fitted to describe LFs are highly correlated between each other. We showed this in the case of our Schechter parameters in \ref{sec:lfmod}. Hence, one by one comparison of these parameters in different studies is challenging. However, the integral of the LF is much more stable. The luminosity weighted integral of the LF at a given redshift gives the luminosity density at that redshift which in turn, once corrected for reddening, provides the star formation rate density at that cosmic age -- an important measurement to understand galaxy evolution over cosmic times. An attempt to derive the extinction of our galaxies is beyond the scope of this paper. Hence, we settled for deriving the luminosity density, a quantity directly derivable from our LFs:

\begin{equation}
LD = \int _{L_{\mathrm{faint}}}^{L_{\mathrm{bright}}}\Phi (L)L \mathrm{d}L \; [\mathrm{Watt}\, \mathrm{Hz}^{-1} \mathrm{Mpc}^{-3}].
\end{equation}

\noindent Following \citet{cucciati12}, we set $L_{\mathrm{faint}}$ and $L_{\mathrm{bright}}$ to the luminosities corresponding to $M_{{UV},\mathrm{faint}} = -3.4$ and $M_{{UV},\mathrm{bright}} = -28.4$. To take into account the errors in the parameters, and the correlation between them, we calculated the integral for each set of parameters given by \texttt{emcee} walkers (see Eq. [\ref{eq:theta_prob}]). Our LD, and its error, are then given by the median and standard deviation of these individual integrations. The resulting LDs are $\mathrm{LD}=19.28^{+0.15}_{-0.11}$, and $\mathrm{LD}=19.44^{+0.26}_{-0.18}$ for $z\sim 3$ and $z\sim 4$, respectively.

We also collected Schechter parameter information from the literature and, to be consistent, carried out ourselves the integration for these parameters using the same integration limits as for our data. Our recollection of literature Schechter parameters and the corresponding LDs calculated by us are given in Table \ref{tab:litvals}. We plot these together with our LD measurements as a function of redshift in Fig. \ref{fig:ld}. We did not calculate the error bars for the literature LDs as we lacked the information on the correlations of the individual Schechter parameters and a simple propagation of individual uncertainties would probably overestimate the errors. However, the scatter of the LD datapoints at similar redshifts gives an idea about the involved uncertainties.

We see that our LDs are compatible with the previous studies. However, we also remind the reader that the $\alpha$ parameter we derive heavily relies on the prior information from the literature because our data is not tracing the faint end of the LF, and the value of $\alpha$ has a strong influence on the derived LD. If we set the integration limits close to the values sampled by our data, $M_{{UV},\mathrm{faint}} = -22.0$ and $M_{{UV},\mathrm{bright}} = -24.5$, the luminosity densities would get down to $\mathrm{LD}=18.20^{+0.03}_{-0.03}$, and $\mathrm{LD}=18.24^{+0.03}_{-0.03}$, for $z\sim 3$ and $z\sim 4$, respectively.

\section{Results: Halo masses}\label{sec:halom}

In this section we derive the masses of the dark matter halos hosting our galaxies from the bias values derived in Sect. \ref{sec:bias}. We compared the derived host halo mass at $z\sim 4$ to that predicted by abundance matching techniques for the galaxies of corresponding $UV$ brightnesses ($UV$ brightness being directly proportional to the SFR). In addition to gaining information about the host halo masses, this exercise serves to double check the bias values we have derived which are based on an assumption of scale independence and which are higher than in any previous study, preventing direct comparison with the literature.

\subsection{Halo mass predicted from the bias}

In the linear regime, the bias values we derived in Sect. \ref{sec:bias} should directly reflect the bias of the dark matter halos hosting our galaxies. To derive the host halo masses from our bias estimates, we followed the modelling of \citet{basilakos08}, as summarised in Appendix \ref{sec:app1}. The host halo masses derived from Eq. (\ref{eq:halom}) for our three luminosity-bias bins are listed in Table \ref{tab:halom}. We see that our galaxies are hosted by high mass dark matter halos, $M_h\sim 0.5 - 3\times 10^{13}h^{-1} M_{\odot}$, as can be expected considering their high luminosities.

\begin{table}
\caption[]{Host halo masses in $10^{13} h^{-1} M_{\odot}$ derived from our bias measurements in two luminosity bins at $z\sim 3$ and at one luminosity bin at $z\sim 4$.}
\label{tab:halom}
\centering
\def\arraystretch{1.5}
\begin{tabular}{r c c c}
\hline
$z$ & $\langle M_{UV}\rangle$ & $b_v$ & $M_h$\\
\hline
$\sim 3$ & -22.2 & $7.7\pm1.2$ & $1.5^{+0.9}_{-0.6}$\\
$\sim 3$ & -22.9 & $9.2\pm2.0$ & $2.6^{+2.2}_{-1.4}$\\
$\sim 4$ & -22.7 & $9.7\pm1.8$ & $0.5^{+0.3}_{-0.2}$\\
\hline
\end{tabular}
\end{table}
\def\arraystretch{1.0}

\begin{figure}
\centering
\includegraphics{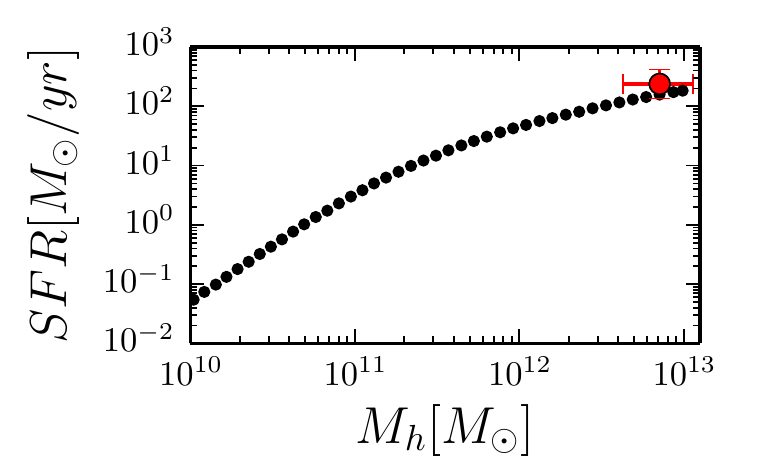}
\caption{SFR as a function of host halo mass. The \citet{mashian16} calibration at $z\sim 4$ is given as a dotted curve. Our bias-based halo mass estimate at $z\sim 4$ and the estimated SFR for the corresponding magnitude bin are given as a red point.}
\label{fig:sfr_halom}
\end{figure}

\subsection{Halo mass predicted from the SFR}

We have derived our bias values assuming that they do not depend on scale. This assumption is necessary as we derived the bias from the cosmic variance, which in turn is a particular case of count--in--cell statistics in which all the scales inside the volume of interest are integrated. However, in reality, the bias is scale dependent. The observational studies at $z\sim 4$ find that the bias is nearly constant for scales larger than $r\sim 0.3h_{70}^{-1}$ Mpc (the linear bias regime), while the bias values become larger at smaller scales \citep{ouchi05}. On the other hand, as discussed in \citet{lopez-sanjuan15}, the bigger is the area covered by the subfields used for the cosmic variance calculation, the smaller is the influence of the non-linear term.

In this section we double check our bias value estimation to assure that the large values that we obtain are not dominated by the non-linear regime. We do this indirectly by deriving the expected host halo mass from the median $UV$ luminosity (assumed to be directly proportional to the SFR) of our galaxies and comparing the result with the bias-based halo mass estimate.

\citet{mashian16} employ abundance matching techniques to calibrate a relation between galaxy SFR and host halo mass by mapping the shape of the observed SFR function at $z=4-8$ to that of the halo mass function. We show the relation given by them at redshift $z\sim 4$ in Fig. \ref{fig:sfr_halom} together with our bias based halo mass estimate at $z\sim 4$ and the corresponding SFR derived from the median $UV$ luminosity at the bin in question. In order to remove the uncertainties related to the conversion from the $M_{UV}$ to SFR, we do the conversion in the same way as was done in \citet{mashian16}: we derive the SFR from the \citet{kennicutt98} relation $\mathrm{SFR}[M_{\odot}yr^{-1}]=1.25\times10^{-28}L_{{UV},corr}[\mathrm{erg}\mathrm{s}^{-1}\mathrm{Hz}^{-1}]$, where $L_{{UV},corr}$ is the ultraviolet luminosity corrected for extinction, and we calculated the $A_{UV}$ from the \citet{meurer99} relation, $A_{1600}=4.43+1.99\beta$, deriving the value of $\beta$ from the relation given by \citet{bouwens14b} at $z\sim 4$, $\beta = -1.85 - 0.11(\mathrm{M}_{UV}+19.5)$. We note that the extinction values and $\beta$ slopes are not intended to be the ideal ones for our data, and are actually different than the ones used in Sect. \ref{sec:ld} above. The idea is simply to strictly follow the \citet{mashian16} prescription in order to remove the systematics related to the conversion from $M_{UV}$ to SFR. For the value of $M_{UV}$ we adopted the median value of the magnitude bin used for the $z\sim 4$ bias calculation in Sec. \ref{sec:cos_var}. The error bars reflect the width of this bin. We applied a small correction, $\Delta M_{UV}=+0.035$ to correct from 1500 \AA~to 1600 \AA~(see Sec. \ref{sec:radio}). Our measurement is consistent, within the error bars, with the $\mathrm{SFR} - M_h$ relation of \citet{mashian16}. 

To conclude, the high host halo mass at $z\sim 4$ derived from the large bias value ($b_v=9.7\pm1.8$) is consistent with the halo mass expected from the luminosity of our galaxies. Hence, we can be confident that the influence of the non-linear regime on our bias values is not significant, and that the bias values we derive, which are larger than in any previous study at these redshifts, are reliable.


\section{Summary and conclusions}\label{sec:conc}

In this work we have calculated the rest frame $UV$ 1500 \AA~ LF at the  redshift range $z=2.5-4.5$ from the data offered by the ALHAMBRA survey using a novel technique based on PDF analysis. We have also estimated the bias values and the corresponding host halo masses for our galaxies. Our main results are summarised as follows:
\begin{itemize}
\item Our LF reveals an excess of bright objects as compared with most of the studies in the literature. However, our LF is compatible with the only magnitude limited spectroscopic LF estimate at these redshifts to date \citep{cucciati12}. Our best Schechter parameters at $z=2.5-3.5$ are $M^{\ast}_{UV}=-21.62\pm 0.11$, $\phi^{\ast}=0.51\pm 0.07\times 10^{-3} \mathrm{Mpc}^3$, $\alpha =-1.53\pm 0.15$, and at $z=3.5-4.5$ are $M^{\ast}_{UV}=-21.67\pm 0.10$, $\phi^{\ast}=0.54\pm 0.08\times 10^{-3} \mathrm{Mpc}^3$, $\alpha =-1.65\pm 0.17$. We note that the $\alpha$s we derive rely heavily on the prior information from the literature. We also derived the luminosity densities at the two redshift bins, giving $\mathrm{log(LD)}=19.28^{+0.15}_{-0.11}$ and $\mathrm{log(LD)}=19.44^{+0.26}_{-0.18}$ at $z\sim 3$ and $z\sim 4$, respectively.
 
\item From the cosmic variance we estimate the bias values for our galaxies. We calculate the bias for two magnitude bins at $z\sim3$, giving $b=7.7\pm 1.2$, and $9.2\pm 2.0$ for $\langle M_{UV} \rangle=-22.2$, and $\langle M_{UV} \rangle =-22.9$, respectively, and at one bin at $z\sim4$, giving $9.7\pm 1.8$ for $\langle M_{UV}\rangle =-22.7$. These bias measurements are tracing brighter galaxies, on average, than in any of the previous studies at these redshifts. Consequently, the derived bias values are higher than in the previous studies.

\item Assuming that the bias values we derived are scale independent, we obtained the host dark matter halo masses corresponding to the measured biases. For the above magnitude bins, in the same order, we derived host halo masses of $1.5^{+0.9}_{-0.6}$, $2.6^{+2.2}_{-1.4}$, and $0.5^{+0.3}_{-0.2}\times 10^{13} h^{-1}M_{\odot}$. We compared the host halo mass derived at $z\sim 4$ to that which would be expected from the SFR corresponding to the median $UV$ magnitude of the bin in question using the $\mathrm{SFR} - M_h$ calibration of \citet{mashian16}. Our bias-based host halo mass estimate is compatible with that expected from our luminosities. Hence, the assumption of scale independence does not strongly affect the bias values we derive.
\end{itemize}

The results of this paper strengthen the evidence -- previously observed in the spectroscopic VVDS data -- of bright end excess in the $UV$ LFs at $z=2.5-4.5$ as compared to previous studies based on broad-band photometry. This bright-end excess can neither be explained by QSO contamination nor by miss-classified low-$z$ galaxies, as the QSO contamination is shown to be insignificant and the large bias values derived for our galaxies would not be compatible with a low-$z$ population. At the faintest bins our LF generally agrees with the literature data, while an agreement is also found at the very brightest bins with the recent wide area study of GOLDRUSH project \citep{ono17} based on broad-band photometry. We call into question the shape of the $z=2.5-4.5$ LF at its bright end; is it a double power-law as suggested by the recent broad-band photometric studies or rather a brighter Schechter function, as suggested by our multi-filter analysis and the spectroscopic VVDS data. Future studies based on the very large area J-PAS \citep[Javalambre Physics of the Accelerating Universe Astrophysical Survey,][]{benitez14b} multi-filter data will, we hope, shed light on this topic.

\begin{acknowledgements}
We acknowledge the anonymous referee for helping us to improve the article. K. Viironen acknowledges the {\it Juan  de  la  Cierva incorporaci\'on} fellowship, IJCI-2014-21960, of the Spanish government. This work has mainly been funded by the FITE (Fondos de Inversiones de Teruel) and the projects AYA2015-66211-C2-1 and AYA2012-30789. We also acknowledge support from the Arag\'on Government Research Group E103 and support from the Spanish Ministry for Economy and Competitiveness and FEDER funds through grants AYA2010-15081, AYA2010-15169, AYA2010-22111-C03-01, AYA2010-22111-C03-02, AYA2011-29517-C03-01, AYA2012-39620, AYA2013-40611-P, AYA2013-42227-P, AYA2013-43188-P, AYA2013-48623-C2-1, AYA2013-48623-C2-2, ESP2013-48274, AYA2014-58861-C3-1, AYA2016-76682-C3-1-P, Generalitat Valenciana projects Prometeo 2009/064 and PROMETEOII/2014/060, Junta de Andaluc\'{\i}a grants TIC114, JA2828, P10-FQM-6444, and Generalitat de Catalunya project SGR-1398. BA has received funding from the European Union's Horizon 2020 research and innovation programme under the Marie Sklodowska-Curie grant agreement No 656354. MP acknowledges financial supports from the Ethiopian Space Science and Technology Institute (ESSTI) under the Ethiopian Ministry of Science Science and Technology (MoST).

This research made use of Astropy, a community-developed core Python package for Astronomy \citet{astropy13}, and Matplotlib, a 2D graphics package used for Python for publication-quality image generation across user interfaces and operating systems \citet{hunter07}. Finally, K. Viironen thanks Isabel, without whom combining child care and research would be too demanding.

\end{acknowledgements}

\bibliographystyle{aa}
\bibliography{/home/CEFCA/kerttu/BIB/oma}

\begin{appendix}

\section{Defining the optimal binning}\label{sec:binning}

\begin{figure}
\centering
\includegraphics{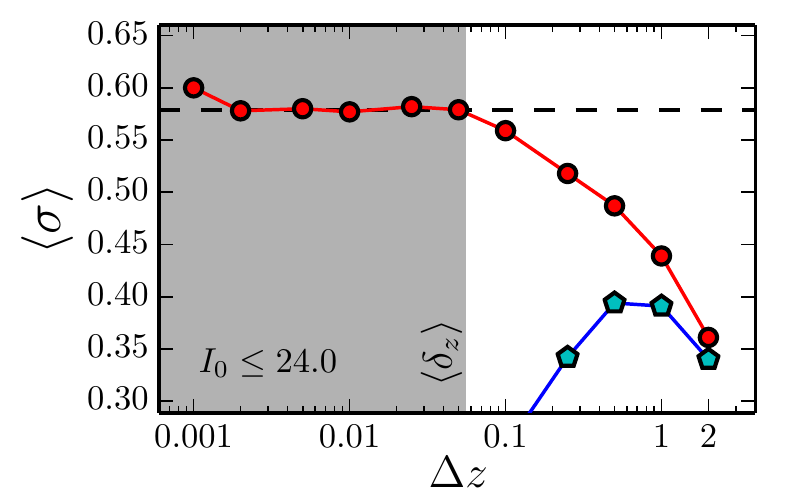}
\caption{Median total variance (red  dots) and the median cosmic variance (cyan pentagons) as a function of the redshift bin size $\Delta z$ at $z=2.5-4.5$. The dashed line marks the total variance in the constant regime. The grey area marks the bin sizes smaller than $\langle \delta z\rangle$, the ALHAMBRA photometric redshift precision.}
\label{fig:bin}
\end{figure}

In this article, we have presented the method to derive the two dimensional, differential, and binned, luminosity and bias functions, and the covariance matrix of the LF. In this Appendix we show how the optimal binning of these function are defined in order, on one hand, to not over-sample the data, and on the other hand, to not lose any information due to too aggressive binning. Once more, we followed the scheme presented in LS17. To ensure a meaningful analysis, we computed the median total and median cosmic variance at $z=2.5-4.5$ within bins of size $\Delta z$, and studied their dependence on the adopted bin size. Both these quantities should decrease if the volume probed by each bin, i.e. the bin size, increases. In Fig. \ref{fig:bin} we see that the total variance starts decreasing at bin sizes close to the  expected redshift uncertainty \citep[$\langle \delta z\rangle = 0.012(1+\langle z\rangle)$,][]{molino14}. At bin sizes smaller than this the signal is dominated by the correlations between adjacent bins and the variance does not change with bin size. However, the cosmic variance only starts decreasing at $\Delta z \gtrsim 1.0$ and we set the redshift bin size to $\Delta z = 1.0$. We note that at bin sizes smaller than $\Delta z \sim 0.5$ the cosmic variance decreases as well. However, we observed that at these small bin sizes the galaxy number counts are no more log-normally distributed, while log-normal distribution should be expected \citep{coles91}, meaning that we are lacking statistics to robustly measure the cosmic variance.

\begin{figure}
\centering
\includegraphics{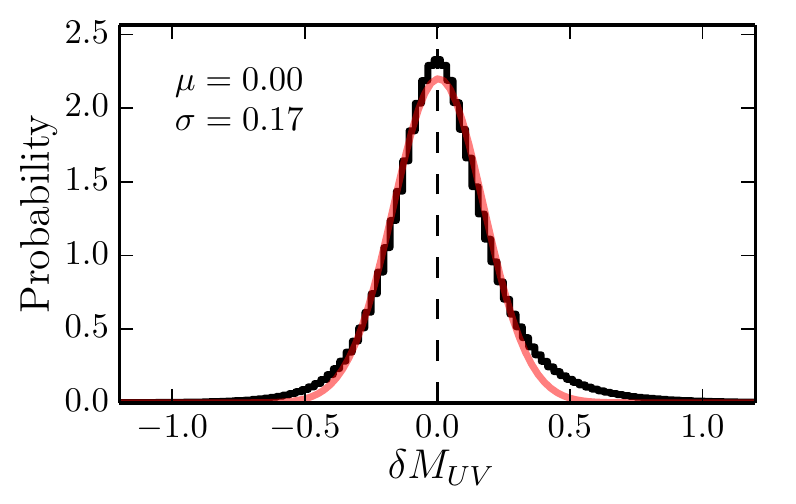}
\caption{Sum of the individual $M_{UV}$ posterior distributions for all the galaxies in our sample normalised to be centred at zero. A Gaussian fit to this distribution is shown as a red line and the median and sigma of this Gaussian are labelled in the panel.}
\label{fig:deltamuv}
\end{figure}

Finally, to study the optimal magnitude bin size, we first created the $M_{UV}$ posterior for each galaxy in our sample::

\begin{equation}
\mathrm{PDF}_i(M_{UV})=\int \mathrm{PDF}_i(z,M_{UV})\mathrm{d}z.
\end{equation}

\noindent Then we summed the individual posterior distributions centred at zero

\begin{equation}
\delta M_{UV}=\sum _i \left [\mathrm{PDF}_i(M_{UV}) - \langle  \mathrm{PDF}_i(M_{UV}\right ) \rangle ],
\end{equation}

\noindent where $\langle \mathrm{PDF}_i(M_{UV}) \rangle$ is the median $UV$ magnitude of the galaxy $i$. The resultant distribution is close to a Gaussian, see Fig. \ref{fig:deltamuv}. We fitted it with a Gaussian function and find a dispersion $\sigma = 0.17$, the mean being zero by definition. Finally, for the LF calculation, we set the magnitude bin size at $\Delta M_{UV} = 0.3$, roughly twice the sigma value of the above distribution. 

To obtain large enough samples to calculate the bias values, much more coarse binning was used; we carried out some tests with the objective of finding the finest possible binning without losing the log-normality of the number count distributions, and ended up calculating bias values at two magnitude bins at $z=2.5-3.5$ and one bias value at $z=3.5-4.5$.

\section{Halo mass from galaxy bias}\label{sec:app1}

In this Appendix we present the compilation of equations from \citet{basilakos08} that we have used to derive the host halo masses in Sect. \ref{sec:halom}. \citet{basilakos08} express the galaxy bias as a function of redshift and halo mass as

\begin{equation}\label{eq:halom}
b(z)=C_1E(z)+C_2E(z)I(z)+1+y_p(z),
\end{equation}

\noindent where 

\begin{equation}
I(z)=\int _z^{\infty} \frac{(1+x)^3}{E^3(x)}\mathrm{d}x,
\end{equation}

\begin{equation}
C_{1,2}(M)\simeq \alpha_{1,2}\left ( \frac{M}{10^{13}h^{-1}M_{\odot}}\right )^{\beta_{1,2}},
\end{equation}

\noindent and $\alpha _1=3.29\pm0.21$, $\beta _1=0.34\pm 0.07$, $\alpha _2=-0.36\pm0.01$, and $\beta _2=0.32\pm 0.06$. The term $y_p(z)$ is defined as

\begin{multline}
y_p(z)=E(z)\int _0^z \frac{f(x)E^2(x)I(x)}{(1+x)^3}\mathrm{d}x \\- E(z)I(z)\int _0^z\frac{f(x)E^2(x)}{(1+x)^3}\mathrm{d}x,
\end{multline}

\noindent where

\begin{equation}
f(z)=A(\nu -2)\frac{(1+z)^{\nu}E(z)}{D(z)},
\end{equation}

\noindent and

\begin{equation}
D(z)=\frac{5\Omega _m E(z)}{2}\int _z^{\infty} \frac{1+x}{E^3(x)}\rm{d}x.
\end{equation}

\noindent We derived the mass dependence of the constant $A$ ourselves by using the values listed in Table 1 of \citet{basilakos08} and fitting a relation $A=a+b\times \log_{10}(M/10^{13}h^{-1}M_{\odot})$. This leads to $a=3.93\times 10^{-3}$ and $b=3.56\times 10^{-3}$. For the constant $\nu$ we adopted the average of its values listed in the same Table 1: $\nu=2.57$. The host halo masses are then obtained from Eq. (\ref{eq:halom}).

\section{Literature compilation of Schechter parameters and the corresponding LDs}

In this Appendix we present Table \ref{tab:litvals} showing a compilation of $UV$ LF Schechter parameters up to $z=6$ from the literature and the LDs that we calculated by integrating the Schechter functions over the range $-28.4 \leq M_{UV} \leq -3.4$.

\begin{table*}
\caption[]{Compilation of literature data for $UV$ LF parameters and the corresponding luminosity densities calculated by us.}
\label{tab:litvals}
\centering
\def\arraystretch{1.4}
\begin{tabular}{l c c c c c}
\hline
Reference & $z$ & $M_{UV}^{\ast}$ & \begin{tabular}{@{}c@{}}$\phi ^\ast$\\ $[$/10$^3$ Mpc$^3]$\end{tabular} & $\alpha$ & \begin{tabular}{@{}c@{}}log(LD)\\ $[$W/Hz/Mpc$^3]$\end{tabular}\\
\hline
Wyder et al. (2005) & 0.05 & $-18.04^{+0.11}_{-0.11}$ & $4.27^{+0.63}_{-0.55}$ & $-1.22^{+0.07}_{-0.07}$ & 18.56\\
Cucciati et al. (2012) & 0.125 & $-18.12$ & $7.00^{+0.44}_{-0.44}$ & $-1.05^{+0.04}_{-0.04}$ & 18.74\\
Cucciati et al. (2012) & 0.300 & $-18.3^{+0.1}_{-0.2}$ & $6.91^{+1.02}_{-0.95}$ & $-1.17^{+0.05}_{-0.05}$ & 18.85\\
Cucciati et al. (2012) & 0.500 & $-18.4^{+0.1}_{-0.1}$ & $6.60^{+0.91}_{-0.86}$ & $-1.07^{+0.07}_{-0.06}$ & 18.84\\
Gabasch et al. (2004) & 0.63 & $-18.17^{+0.11}_{-0.11}$ & $11.0^{+0.7}_{-0.6}$ & $-1.07$ & 18.97\\
Cucciati et al. (2012) & 0.700 & $-18.3^{+0.1}_{-0.1}$ & $9.53$ & $-0.90^{+0.08}_{-0.08}$ & 18.92\\
Oesch et al. (2010) & 0.75 & $-19.17^{+0.51}_{-0.51}$ & $3.02^{+3.15}_{-1.54}$ & $-1.52^{+0.25}_{-0.25}$ & 19.05 \\
Cucciati et al. (2012) & 0.9 & $-18.7^{+0.1}_{-0.1}$ & $9.01^{+0.94}_{-0.96}$ & $-0.85^{+0.10}_{-0.10}$ & 19.04\\
Gabasch et al. (2004) & 0.96 & $-18.85^{+0.10}_{-0.10}$ & $10.3^{+0.6}_{-0.6}$ & $-1.07$ & 19.21\\
Cucciati et al. (2012) & 1.100 & $-19.0^{+0.2}_{-0.2}$ & $7.43^{+1.08}_{-1.15}$ & $-0.91^{+0.16}_{-0.16}$ & 19.09\\
Oesch et al. (2010) & 1.25 & $-20.08^{+0.36}_{-0.36}$ & $1.26^{+0.98}_{-0.55}$ & $-1.84^{+0.15}_{-0.15}$ & 19.49\\
Gabasch et al. (2004) & 1.36 & $-19.48^{+0.11}_{-0.11}$ & $5.6^{+0.6}_{-0.5}$ & $-1.07$ & 19.20\\
Cucciati et al. (2012) & 1.45 & $-19.6^{+0.2}_{-0.2}$ & $4.1^{+0.77}_{-0.87}$ & $-1.09^{+0.23}_{-0.23}$ & 19.12\\
Oesch et al. (2010) & 1.5 & $-19.82^{+0.51}_{-0.51}$ & $2.29^{+2.61}_{-1.22}$ & $-1.46^{+0.54}_{-0.54}$ & 19.14\\
Sawicky \& Thompson (2006) & 1.7 & $-19.80^{+0.32}_{-0.26}$ & $16.98^{+4.90}_{-3.80}$ & $-0.81^{+0.21}_{-0.15}$ & 19.75 \\
Parsa et al. (2016) & 1.7 & $−19.61^{+0.07}_{-0.07}$ & $6.81^{+0.81}_{-0.81}$ & $-1.33^{+0.03}_{-0.03}$ & 19.44\\
Oesch et al. (2010) & 1.75 & $-20.17^{+0.34}_{-0.34}$ & $2.34^{+1.64}_{-0.96}$ & $-1.60^{+0.21}_{-0.21}$ & 19.42\\
Gabasch et al. (2004) & 1.88 & $-19.97^{+0.22}_{-0.24}$ & $3.3^{+0.6}_{-0.6}$ & $-1.07$ & 19.16\\
Oesch et al. (2010) & 1.9 & $-20.16^{+0.52}_{-0.52}$ & $2.19^{+2.82}_{-1.23}$ & $-1.60^{+0.51}_{-0.51}$ & 19.39 \\
Parsa et al. (2016) & 1.9 & $-19.68^{+0.05}_{-0.05}$ & $7.02^{+0.66}_{-0.66}$ & $-1.32^{+0.03}_{-0.03}$ & 19.48\\
Cucciati et al. (2012) & 2.1 & $-20.4^{+0.1}_{-0.1}$ & $3.37^{+0.24}_{-0.24}$ & $-1.30$ & 19.44\\
Sawicky \& Thompson (2006) & 2.2 & $-20.60^{+0.38}_{-0.44}$ & $3.02^{+1.77}_{-1.36}$ & $-1.20^{+0.24}_{-0.22}$ & 19.42\\
Parsa et al. (2016) & 2.25 & $-19.71^{+0.07}_{-0.07}$ & $7.59^{+0.88}_{-0.88}$ & $-1.26^{+0.04}_{-0.04}$ & 19.50\\
Parsa et al. (2016) & 2.25 & $-19.99^{+0.08}_{-0.08}$ & $6.2^{+0.77}_{-0.77}$ & $-1.31^{+0.04}_{-0.04}$ & 19.54\\
Oesch et al. (2010) & 2.50 & $-20.69^{+0.17}_{-0.17}$ & $3.24^{+1.03}_{-0.78}$ & $-1.73^{+0.11}_{-0.11}$ & 19.94\\
Gabasch et al. (2004) & 2.53 & $-20.61^{+0.09}_{-0.09}$ & $3.2^{+0.2}_{-0.2}$ & $-1.07$ & 19.41\\
Parsa et al. (2016) & 2.8 & $-20.20^{+0.07}_{-0.07}$ & $5.32^{+0.6}_{-0.6}$ & $-1.31^{+0.04}_{-0.04}$ & 19.56 \\
Sawicky \& Thompson (2006) & 3.0 & $-20.90^{+0.22}_{-0.14}$ & $1.7^{+0.59}_{-0.32}$ & $-1.43^{+0.17}_{-0.09}$ & 19.42\\
Cucciati et al. (2012) & 3.0 & $-21.4^{+0.1}_{-0.1}$ & $0.86^{+0.05}_{-0.05}$ & $-1.50$ & 19.38\\
Bouwens et al. (2015) & 3.0 & $-20.97^{+0.14}_{-0.14}$ & $1.71^{+0.53}_{-0.53}$ & $-1.73^{+0.13}_{-0.13}$ & 19.78\\
{\bf This work} & 3.0 & $-21.62\pm0.11$ & $0.51\pm0.07$ & $-1.53\pm0.15$ & $19.28^{+0.15}_{-0.11}$\\
Gabasch et al. (2004) & 3.46 & $-20.72^{+0.09}_{-0.10}$ & $2.3^{+0.2}_{-0.2}$ & $-1.07$ & 19.31\\
Paltani et al. (2007) & 3.5 & $-21.49^{+0.19}_{-0.19}$ & $1.24^{+0.48}_{-0.5}$ & $-1.4$ & 19.50\\
Bouwens et al. (2007) & 3.8 & $-21.06^{+0.10}_{-0.10}$ & $1.1^{+0.2}_{-0.2}$ & $-1.76^{+0.05}_{-0.05}$ & 19.67\\
Bouwens et al. (2015) & 3.8 & $-20.88^{+0.08}_{-0.08}$ & $1.97^{+0.34}_{-0.29}$ & $-1.64^{+0.04}_{-0.04}$ & 19.68\\
Bouwens et al. (2015) & 3.8 & $-20.88^{+0.08}_{-0.08}$ & $1.97^{+0.34}_{-0.29}$ & $-1.64^{+0.04}_{-0.04}$ & 19.68\\
Parsa et al. (2016) & 3.8 & $-20.71^{+0.10}_{-0.10}$ & $2.06^{+0.33}_{-0.33}$ & $-1.43^{+0.04}_{-0.04}$ & 19.43\\
Sawicky \& Thompson (2006) & 4.0 & $-21.00^{+0.40}_{-0.46}$ & $0.85^{+0.53}_{-0.45}$ & $-1.26^{+0.40}_{-0.36}$ & 19.06\\
Cucciati et al. (2012) & 4.0 & $-22.2^{+0.2}_{-0.2}$ & $0.11^{+0.01}_{-0.01}$ & $-1.73$ & 19.08\\
{\bf This work} & 4.0 & $-21.67\pm0.10$ & $0.54\pm0.08$ & $-1.65\pm0.17$ & $19.44^{+0.26}_{-0.18}$\\
Gabasch et al. (2004) & 4.51 & $-21.00^{+0.15}_{-0.11}$ & $1.0^{+0.1}_{-0.1}$ & $-1.07$ & 19.06\\
Bouwens et al. (2015) & 4.9 & $-21.10^{+0.15}_{-0.15}$ & $0.79^{+0.23}_{-0.18}$ & $-1.76^{+0.06}_{-0.06}$ & 19.54\\
Bouwens et al. (2015) & 4.9 & $-21.17^{+0.12}_{-0.12}$ & $0.74^{+0.18}_{-0.14}$ & $-1.76^{+0.05}_{-0.05}$ & 19.54\\
Bouwens et al. (2007) & 5.0 & $-20.69^{+0.13}_{-0.13}$ & $0.9^{+0.3}_{-0.2}$ & $-1.69^{+0.09}_{-0.09}$ & 19.33\\
Iwata et al. (2007) & 5.0 & $-21.28^{+0.38}_{-0.38}$ & $0.41^{+0.29}_{-0.3}$ & $-1.48^{+0.38}_{-0.32}$ & 18.99\\
Mclure et al. (2009) & 5.0 & $-20.73^{+0.11}_{-0.11}$ & $0.9^{+0.2}_{-0.2}$ & $-1.66^{+0.06}_{-0.06}$ & 19.30\\
Bouwens et al. (2007) & 5.9 & $-20.29^{+0.19}_{-0.19}$ & $1.2^{+0.6}_{-0.4}$ & $-1.77^{+0.16}_{-0.16}$ & 19.42\\
Bouwens et al. (2015) & 5.9 & $-20.94^{+0.20}_{-0.20}$ & $0.50^{+0.22}_{-0.16}$ & $-1.87^{+0.10}_{-0.10}$ & 19.51\\
Bouwens et al. (2015) & 5.9 & $-21.10^{+0.24}_{-0.24}$ & $0.39^{+0.21}_{-0.14}$ & $-1.90^{+0.10}_{-0.10}$ & 19.55\\
Mclure et al. (2009) & 6.0 & $-20.04^{+0.12}_{-0.12}$ & $1.8^{+0.5}_{-0.5}$ & $-1.71^{+0.11}_{-0.11}$ & 19.39 \\
\hline
\end{tabular}
\end{table*}
\def\arraystretch{1}
\end{appendix}

\end{document}